\documentclass[11pt]{article}
\usepackage[utf8]{inputenc}

\title{Dynamic Bayesian Networks with Conditional Dynamics in Edge Addition and Deletion}
\author{Lupe S. H. Chan \hspace{.2cm}\\
	The Hong Kong University of Science and Technology\\
	and \\
	Amanda M. Y. Chu \\
	The Education University of Hong Kong\\
	and \\
	Mike K. P. So \thanks{Corresponding author}\\
	The Hong Kong University of Science and Technology}
\date{\today}

\usepackage{setspace}
\usepackage[a4paper]{geometry}
\usepackage{hyperref}
\usepackage{physics}
\usepackage{amsmath,amssymb}
\usepackage{mathtools}
 \usepackage[table,xcdraw]{xcolor}
\usepackage[normalem]{ulem}

\usepackage{tikz}
\usepackage{float}

\usepackage{color,soul}
\definecolor{myGreen}{rgb}{0.8,1,0.2}
\definecolor{myRed}{rgb}{0.9,0.65,0.65}

\usepackage{enumerate}
\usetikzlibrary{bayesnet}
\usetikzlibrary{shapes.geometric}%

    \usepackage{tikz}
    \usepackage{graphicx}
     \usepackage{array,multirow,graphicx}

\usepackage{natbib}
\usepackage{tikz}
\usepackage{float}
\usepackage{caption}
\usepackage{subcaption}
\usepackage{bbm}
\usetikzlibrary{calc}
\usepackage{siunitx}
\usetikzlibrary{bayesnet}
\usetikzlibrary{shapes.geometric}%
\usetikzlibrary{shadows,arrows.meta,positioning,backgrounds,fit}
\usetikzlibrary{arrows,decorations.pathreplacing}

\tikzset{%
  materia/.style={draw, fill=blue!10, text width=28em,text centered, minimum height=1.5em},
  cont/.style={draw, fill=yellow!10, text width=28em,text centered, minimum height=1.5em},
   dist/.style={draw, fill=red!10, text width=28em,text centered, minimum height=1.5em},
      mix/.style={draw, fill=green!10, text width=28em,text centered, minimum height=1.5em},
    etape/.style={materia, text width=13em, minimum width=13em, minimum height=3em, rounded corners}, 
    wideEtape/.style={materia, text width=12em, minimum width=12em, minimum height=3em, rounded corners}, 
    contEtape/.style={cont, text width=12em, minimum width=12em, minimum height=3em, rounded corners},
        mixEtape/.style={mix, text width=12em, minimum width=12em, minimum height=3em, rounded corners},
        distEtape/.style={dist, text width=12em, minimum width=12em, minimum height=3em, rounded corners},
  texto/.style={above, text width=6em, text centered},
  linepart/.style={draw, thick, color=black!50, -LaTeX, dashed},
  line/.style={draw, thick, color=black!50, -LaTeX},
  ur/.style={draw, text centered, minimum height=0.01em,draw=white!100,text width=13em},
  result/.style={draw, text centered, minimum height=0.01em,draw=white!100,text width=17em},
  back group/.style={fill=yellow!10,rounded corners, draw=black!50, dashed, inner xsep=15pt, inner ysep=10pt},
     brace_top/.style={
     color=blue,
     decoration={brace},
     decorate
   },
   brace_bottom/.style={
     color=blue,
     decoration={brace, mirror},
     decorate
   },number line/.style={}
}

\usepackage{algpseudocode}

\newcommand\bo[1]{\boldsymbol{#1}}

\usepackage{amsthm}
\theoremstyle{definition}

\newcommand\numberthis{\addtocounter{equation}{1}\tag{\theequation}}

\begin{document}
		 \maketitle
		 	\thispagestyle{empty}
	\begin{abstract}
		This study presents a dynamic Bayesian network framework that facilitates intuitive gradual edge changes. We use two conditional dynamics to model the edge addition and deletion, and edge selection separately. Unlike previous research that uses a mixture network approach, which restricts the number of possible edge changes, or structural priors to induce gradual changes, which can lead to unclear network evolution, our model induces more frequent and intuitive edge change dynamics. We employ Markov chain Monte Carlo (MCMC) sampling to estimate the model structures and parameters and demonstrate the model's effectiveness in a portfolio selection application.
	\end{abstract}

\section{Introduction}
Dynamic Bayesian networks (DBNs) have been used to model the conditional dependencies in time series data. DBNs have been used widely in various fields, for example, in finance and economics \citep{ammann2007prior,jangmin2004stock,wang2017stock,wang2020fmdbn},  biomedical science \citep{VANDERHEIJDEN201494,van2008dynamic,marini2015dynamic,zandona2019dynamic} and bioinformatics \citep{dondelinger2010heterogeneous,Robinson2010,Grzegorczyk2009,murphy1999modelling,grzegorczyk2008modelling,jia2010constructing}. In many applications, especially in the applications in finance and economics, as well as biomedical science, homogenous Markov process is commonly assumed in the DBN. This means that the structure linking variables from time $t-1$ to $t$ remains unchanged for different $t$. However, DBNs have seen greater development in bioinformatics. Several attempts are made to extend the DBNs to the non-stationary domain; \cite{murphy1999modelling} introduces a DBN model in which two consecutive networks are linked and regularized using a regression-like structure; \cite{grzegorczyk2008modelling,Grzegorczyk2009} introduce a network mixture model, which allows the networks and parameters to vary in different time segments from the mixture of components; \cite{Robinson2010} introduces a non-stationary DBN model, where the networks vary in different segments, and the number of edge changes is penalized using an exponential prior;
More examples using segmentwise ideas and structure regularizations are in \cite{wang2010time,jia2010constructing,dondelinger2010heterogeneous}.

A practical problem is that, whereas network mixture models allow networks to vary in different segments, as the parameters are estimated within segments, the number of segments cannot be too large to ensure that the sample sizes are not too small in each of the segments. This restricts the number of possible edge changes. Conceptually, the time series of networks should vary slowly over time; for financial networks,  dependence structures among financial securities are likely to vary over time; for clinical networks, the dependence structures among the clinical indicators could change over time once interventions on the patients started. Thus, network mixture models are unrealistic. To allow the network to vary slowly over time, structural priors are used in literature to produce smooth time series of networks. This solves the practical problem of the mixture models, however, the evolution of the networks remains unclear.

We present a dynamic Bayesian network model with conditional dynamics in edge addition and deletion (DBN-AD) to introduce intuitive gradual edge changes in DBNs. An evolution step of a network at time $t-1$ to $t$ is divided into two substeps: (i) Determining the numbers of edge addition and deletion, and (ii) edge selection. The numbers of edge addition and deletion are respectively modeled by two dynamics inspired by the generalized autoregressive conditional heteroskedasticity (GARCH) model \citep{engle1982autoregressive,bollerslev1986generalized} and the autoregressive conditional Poisson model \citep{Heinen2003}. After determining these numbers, we select edges to remove from and add to the network at time $t-1$ based on the activeness of the nodes. The activeness of a node measures the willingness of the node to make a change. It follows a logistic model with exogenous factors.

In Section \ref{section:methodology}, we provide the methodology of the proposed model. In Section \ref{section:simulation_study}, we present the results of the simulation study to demonstrate that the parameters and networks can be recovered with good accuracy from the simulated data sets. In Section \ref{section:empirical_fin_app_DBN}, we showcase the superiority of our model in an empirical study of portfolio selection. In Section \ref{section:conclusion}, we give a discussion and a conclusion of this chapter.

\section{Methodology}
\label{section:methodology}
\subsection{Conditional Addition and Deletion Dynamics}
\label{section:condition_add_and_delete_dynamics}
Let $\mathbf{X}_t=(X_{1t},\ldots,X_{nt})^T$ for $t=1,\ldots,T$ be a vector time series. The dependence structure at time $t$ is encoded by a directed acyclic graph (DAG) $G_t$. We assume that $G_t$ is evolved from $G_{t-1}$ by adding $a_t$ edges and removing $d_t$ edges, where $a_t$ and $d_t$ are two non-negative integer-valued random variables. We assume that $a_t$ and $d_t$ have the following dynamics for $t\geq 2$:
\begin{equation}
	\begin{aligned}
		a_t &\sim \text{Poisson}(\mu_t^a),~d_t\sim \text{Poisson}(\mu_t^d),\\
		\log \mu_t^a &= (1-\alpha_1-\beta_1)\log \overline{\mu}^a + \alpha_1 \log \mu_{t-1}^a + \beta_1 a_{t-1} + g_a(\mathbf{v}_{t-1}),\\
		\log \mu_t^d &= (1-\alpha_2-\beta_2)\log \overline{\mu}^d + \alpha_2 \log \mu_{t-1}^d + \beta_2 d_{t-1} + g_d(\mathbf{v}_{t-1}),
	\end{aligned}
	\label{eqt:dynamic_mean_spec}
\end{equation}
where $\mathbf{v}_{t-1}$ is a vector of exogenous variables at time $t-1$ driving the dynamics in \eqref{eqt:dynamic_mean_spec}, and $g_a$ and $g_b$ are two functions mapping $\mathbf{v}_{t-1}$ to real numbers. For example, $g_a(\mathbf{v}_{t-1})=\boldsymbol{\gamma}_a^T \mathbf{v}_{t-1}$ and $g_d(\mathbf{v}_{t-1})=\boldsymbol{\gamma}_d^T \mathbf{v}_{t-1}$ with two vectors of coefficients $\boldsymbol{\gamma}_a$ and $\boldsymbol{\gamma}_d$. $ \overline{\mu}^a, \overline{\mu}^d >0$ are parameters in the intercepts, $\alpha_1,\alpha_2,\beta_1$ and $\beta_2$ are persistence parameters with constraints $\alpha_1,\alpha_2,\beta_1,\beta_2\geq 0$ and $\alpha_1+\beta_1,\alpha_2+\beta_2<1$. For $t=1$, we initialize $\mu_1^a=\overline{\mu}^a$ and $\mu_1^d=\overline{\mu}^d$.

\subsection{Edge Selection Dynamics}
\label{section:edge_selection_dynamics}
After $a_t$ and $d_t$ are generated from \eqref{eqt:dynamic_mean_spec}, we need to decide which edges to add and delete. The selection is based on a list of activenesses of the edges. The activeness of a pair $(i,j)$ at time $t$ is measured by $w_{ij,t}\in (0,1)$, defined as
\begin{equation}
	w_{ij,t}=w_{i,t}\cdot w_{j,t},
	\label{eqt:wijEqt}
\end{equation}
where $w_{i,t}$ measures the individual activeness of node $i$ at time $t$. A larger $w_{ij,t}$ of an unconnected pair indicates that the edges $i\to j$ and $j\to i$ have higher priorities to be added to the network. On the other hand, a small $w_{ij,t}$ of an existing edge in a network indicates that the edge has a higher priority to be removed. The product between $w_{i,t}$ and $w_{j,t}$ in \eqref{eqt:wijEqt} can be viewed as the collaboration between nodes $i$ and $j$, such that $w_{ij,t}$ is large when both nodes $i$ and $j$ have high levels of activeness. We model the individual activeness $w_{i,t}$ using an exponentially weighted moving average (EWMA) model:
\begin{equation}
	\ln\left(\frac{w_{i,t}}{1-w_{i,t}}\right) = (1-\beta^{ES}) \ln\left(\frac{w_{i,t-1}}{1-w_{i,t-1}}\right) +  \beta_{ES}\cdot g_{ES}(\mathbf{x}_{i,t-1}),
	\label{eqt:dynamics_wit_reference_group}
\end{equation}
where $\mathbf{x}_{i,t-1}$ is a vector of exogenous variables driving the dynamics in \eqref{eqt:dynamics_wit_reference_group}, $g_{ES}$ is a function mapping $\mathbf{x}_{i,t-1}$ to a real number, and $\beta^{ES}\in (0,1)$ is a parameter. The logit transformation in \eqref{eqt:dynamics_wit_reference_group} is to ensure that $w_{i,t}\in (0,1)$.

The model starts at the deletion phase. The edges that are connected in $G_{t-1}$ are ordered ascendingly according to $w_{ij,t}$ to form a deletion list. The $d_t$ edges with the smallest $w_{ij,t}$ in the deletion list are removed from $G_{t-1}$. The model then proceeds to the addition phase. The edges that are not connected in $G_{t-1}$ are ordered descendingly according to $w_{ij,t}$ to form an addition list. The $a_t$ edges with the largest $w_{ij,t}$ that result in DAGs in the addition list are added to the network one by one. Every time an edge is added, the model checks again if adding the remaining edges results in DAGs. Note that, since edges $i\to j$ and $j\to i$ can co-exist in the addition list, they will have the same $w_{ij,t}$. To ensure the uniqueness of the addition list, we order $i\to j$ first if $w_{i,t}>w_{j,t}$, and order $j\to i$ first otherwise. Then, we assume that $w_{ij,t}$ should take distinct values for different $i$ and $j$ so that the edges can be uniquely ordered. In practice, $w_{ij,t}$'s are distinct almost all the time since $w_{ij,t}$'s take continuous values.

\begin{figure}[H]
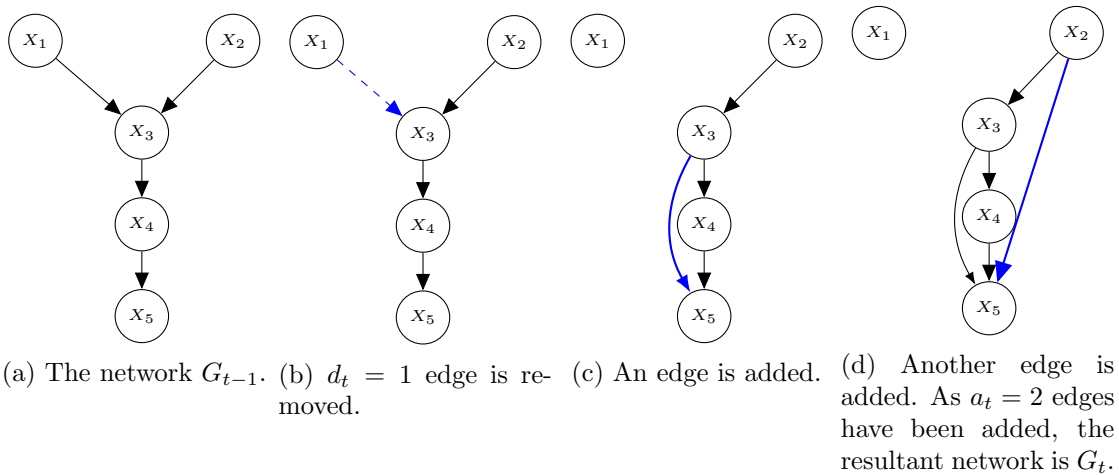

	\begin{subfigure}{0.245\textwidth}
		\centering
		\tikz{
			\node[latent,xshift=-1.3cm] (1) {\tiny $X_{1}$};%
			\node[latent,xshift=1.3cm] (2) {\tiny $X_{2}$}; %
			\node[latent,below=0.5 cm of 1,xshift=1.4cm] (3) {\tiny $X_{3}$}; 
			\node[latent,below=0.5 cm  of 3] (4) {\tiny $X_{4}$};%
			\node[latent,below=0.5 cm  of 4] (5) {\tiny $X_{5}$};%
			\edge {1} {3}
			\edge{2}{3}
			\edge {3} {4}
			\edge {4} {5}
		}
		\caption{The network $G_{t-1}$.\\[1.2cm]}
		\label{fig:network_eg_step1}
	\end{subfigure}
	\begin{subfigure}{0.245\textwidth}
		\centering
		\tikz{
			\node[latent,xshift=-1.3cm] (1) {\tiny$X_{1}$};%
			\node[latent,xshift=1.3cm] (2) {\tiny $X_{2}$}; %
			\node[latent,below=0.5 cm of 1,xshift=1.4cm] (3) {\tiny $X_{3}$}; 
			\node[latent,below=0.5 cm of 3] (4) {\tiny $X_{4}$};%
			\node[latent,below=0.5 cm of 4] (5) {\tiny $X_{5}$};%
			\edge [dashed,blue]{1} {3}
			\edge {3} {4}
			\edge {4} {5}
			\edge{2}{3}
		}
		\caption{$d_{t}=1$ edge is removed.\\[0.3cm]}
		\label{fig:network_eg_step2}
	\end{subfigure}
	\begin{subfigure}{0.245\textwidth}
		\centering
		\tikz{
			\node[latent,xshift=-1.3cm] (1) {\tiny$X_{1}$};%
			\node[latent,xshift=1.3cm] (2) {\tiny $X_{2}$}; %
			\node[latent,below=0.5 cm of 1,xshift=1.4cm] (3) {\tiny $X_{3}$}; 
			\node[latent,below=0.5 cm of 3] (4) {\tiny $X_{4}$};%
			\node[latent,below=0.5 cm of 4] (5) {\tiny $X_{5}$};%
			\edge {2} {3}
			\edge {3} {4}
			\edge {4} {5}
			\path[->,-{Latex}] (3) edge[bend right,thick,blue] node [right] {} (5);
		}
		\caption{An edge is added.\\[1.2cm]}
		\label{fig:network_eg_step3}
	\end{subfigure}
	\begin{subfigure}{0.245\textwidth}
		\centering
		\tikz{
			\node[latent,xshift=-1.3cm] (1) {\tiny$X_{1}$};%
			\node[latent,xshift=1.3cm] (2) {\tiny $X_{2}$}; %
			\node[latent,below=0.5 cm of 1,xshift=1.45cm] (3) {\tiny $X_{3}$}; 
			\node[latent,below=0.5 cm of 3] (4) {\tiny $X_{4}$};%
			\node[latent,below=0.5 cm of 4] (5) {\tiny $X_{5}$};%
			\edge {2} {3}
			\edge {3} {4}
			\edge {4} {5}
			\edge[thick,blue] {2} {5}
			\path[->,-{Latex}] (3) edge[bend right] node [right] {} (5);
		}
		\caption{Another edge is added. As $a_{t}=2$ edges have been added, the resultant network is $G_t$.}
		\label{fig:network_eg_step4}
	\end{subfigure}
	\caption{Illustration of how $G_{t-1}$ evolves into $G_t$ with $d_t=1$ and $a_t=2$ in the DBN-AD model. The blue dashed line indicates the removed edge and the blue solid line in each step indicates the newly added edge. $X_i$'s in the figures indicate the $i$th variable in the network, for $i=1,2,3,4,5$.}
	\label{fig:example_3_stepwise_new}
\end{figure}

To illustrate the deletion and addition phases, we show an example with $d_t=1$ and $a_t=2$. $G_{t-1}$, the network at time $t-1$, is shown in \autoref{fig:network_eg_step1}. Suppose that $w_{1,t}=0.1$, $w_{2,t}=0.2$, $w_{3,t}=0.3$, $w_{4,t}=0.4$, and $w_{5,t}=0.5$. \autoref{tab:deletion_list_example} shows the deletion list. As $d_{t}=1$, the edge with the smallest $w_{ij,t}$, i.e., $1\to 3$, is removed. The deletion phase ends since $d_t=1$ edge has been removed, and the resultant network is shown in \autoref{fig:network_eg_step2}, where the blue dashed edge indicates that the edge has been removed. The model then proceeds to the addition phase. \autoref{tab:addition_list_example1} shows the addition list right after the deletion phase. Note that $1\to 3$ is not in the addition list as we do not allow the addition phase to add back the edges removed in the deletion phase. Furthermore, only edges resulting in DAGs can be added. The edge with the largest $w_{ij,t}$ that results in a DAG is first added, i.e., $3\to 5$. The resultant network after adding the edge is shown in \autoref{fig:network_eg_step3}, where the blue solid edge indicates the new edge. The model checks again if adding the remaining edges in the addition list shown in \autoref{tab:addition_list_example2} will result in DAGs. The edge with the second largest $w_{ij,t}$ and results in a DAG, i.e., $2\to 5$, is added to form the network in \autoref{fig:network_eg_step4}. As $a_t=2$ edges have been added, the addition phase ends, and $G_{t-1}$ evolves into $G_t$ in \autoref{fig:network_eg_step4}.

\begin{table}[H]
	\centering 
	\begin{tabular}{lllllll}
		$i$          & $j$          & $w_{i,t}$ & $w_{j,t}$ & $w_{ij,t}$ & Action                \\ \hline
		1          & 3          & 0.1       & 0.3       & 0.03       & Delete    \\
		{2} & {3} & 0.2       & 0.3       & 0.06       & -       \\
		3          & 4          & 0.3       & 0.4       & 0.12       & -           \\
		4          & 5          & 0.4       & 0.5       & 0.2        & -          
	\end{tabular}
	\caption{An illustration of the deletion list and the deletion process. Each entry under ``Action'' indicates if the edge will be deleted from the network. The edge $1 \to 3$ is removed.}
	\label{tab:deletion_list_example}
\end{table}

\begin{table}[H]
	\centering 
	\begin{subtable}{0.45\textwidth}
		\centering \scriptsize
		\begin{tabular}{lllllll}
			$i$ & $j$ & $w_{i,t}$ & $w_{j,t}$ & $w_{ij,t}$ & DAG & Action \\ \hline
			
			5 & 3 & 0.5       & 0.3       & 0.15       & No     &  -     \\
			3 & 5 & 0.3       & 0.5       & 0.15       & Yes     & Add     \\
			5 & 2 & 0.5       & 0.2       & 0.1        & No     & -       \\
			2 & 5 & 0.2       & 0.5       & 0.1        & Yes     & -           \\
			4 & 2 & 0.4       & 0.2       & 0.08       & No     & -           \\
			2 & 4 & 0.2       & 0.4       & 0.08       & Yes     & -           \\
			5 & 1 & 0.5       & 0.1       & 0.05       & Yes      & -           \\
			1 & 5 & 0.1       & 0.5       & 0.05       & Yes     & -           \\
			4 & 1 & 0.4       & 0.1       & 0.04       & Yes      & -           \\
			1 & 4 & 0.1       & 0.4       & 0.04       & Yes     & -           \\
			3 & 1 & 0.3 	  & 0.1       & 0.03       & Yes     & -        \\
			2 & 1 & 0.2       & 0.1       & 0.02       & Yes     & -           \\
			1 & 2 & 0.1       & 0.2       & 0.02       & Yes     & -          
		\end{tabular}
		\caption{The edge $3 \to 5$ is added.\\[0.9cm]}
		\label{tab:addition_list_example1}
	\end{subtable}
	\hspace{3mm}
	\begin{subtable}{0.45\textwidth}
		\centering \scriptsize
		\begin{tabular}{lllllll}
			$i$ & $j$ & $w_{i,t}$ & $w_{j,t}$ & $w_{ij,t}$ & DAG     & Action \\ \hline
			\rowcolor[HTML]{BFBFBF} 
			5 & 3 &           &           &            &             &             \\
			\rowcolor[HTML]{BFBFBF} 
			3 & 5 &           &           &            &             &             \\
			5 & 2 & 0.5       & 0.2       & 0.1        & No     & -       \\
			2 & 5 & 0.2       & 0.5       & 0.1        & Yes     & Add           \\
			4 & 2 & 0.4       & 0.2       & 0.08       & No         & -       \\
			2 & 4 & 0.2       & 0.4       & 0.08       & Yes& -           \\
			5 & 1 & 0.5       & 0.1       & 0.05       & Yes          & -           \\
			1 & 5 & 0.1       & 0.5       & 0.05       & Yes         & -           \\
			4 & 1 & 0.4       & 0.1       & 0.04       & Yes          & -           \\
			1 & 4 & 0.1       & 0.4       & 0.04       & Yes         & -           \\
			3 & 1 & 0.3 	  & 0.1       & 0.03       & Yes     & -        \\
			2 & 1 & 0.2       & 0.1       & 0.02       & Yes & -           \\
			1 & 2 & 0.1       & 0.2       & 0.02       & Yes         & -           
		\end{tabular}
		\caption{The updated addition list with DAG rechecking. The edge $2 \to 5$ is added to $G_{t-1}$.}
		\label{tab:addition_list_example2}
	\end{subtable}
	\caption{An illustration of the addition list and the edge addition process. Each entry under ``DAG'' checks if the resultant network is a DAG after adding the edge on the same row, and each entry under ``Action'' indicates if the edge will be added to the network.}
	\label{tab:addition_list_example}
\end{table}

\subsection{Volatility and Conditional Correlation Dynamics}
\label{section:volatility_and_correlation}
We will showcase our model in a financial application. In a portfolio selection problem, $\mathbf{X}_t$ is a vector of the daily log-returns of a portfolio of $n$ stocks. The dependence structure of $\mathbf{X}_t$ at time $t$ is encoded by a DAG $G_t$ and the covariance matrix $\Sigma_{t}$. At each time $t$, $G_{t-1}$ and $\Sigma_{t-1}$ are known, and the network $G_{t-1}$ evolves into $G_t$ according to the methodology in Sections \ref{section:condition_add_and_delete_dynamics} and \ref{section:edge_selection_dynamics}. The conditional covariance matrix at time $t-1$, $\Sigma_{t-1}$, also evolves into $\Sigma_{t}$ according to the dependence structure in $G_t$. Inspired by the Dynamic Conditional Correlation (DCC)-GARCH model \citep{engle2002dynamic}, we model the marginal distribution of $X_{it}$ with the GARCH process
\begin{equation}
	\begin{aligned}
		X_{it} &= \sigma_{it}\varepsilon_{it},\\
		\sigma_{it}^2 &= (1-\alpha_{i}^G-\beta_{i}^G)\overline{\sigma}_i^2+ \alpha_{i}^{G}X_{i,t-1}^2 + \beta_{i}^{G}\sigma_{i,t-1}^2,
	\end{aligned}
	\label{eqt:GARCH}
\end{equation}
where $\bo{\varepsilon}_{t}=(\varepsilon_{1t},\ldots,\varepsilon_{nt})$ follows a multivariate normal distribution with the mean of zero and correlation matrix $R_t$,  $\bo{\varepsilon}_{s}$ and $\bo{\varepsilon}_{t}$ are independent for $s\ne t$, $\overline{\sigma}_i^2$ is the long-run variance and $\alpha_{i}^G$ and $\beta_{i}^G$ are the persistence parameters for the $i$th stock, $i=1,\ldots,n$. {In general, other GARCH models can be used. For example, we can use the exponential GARCH (EGARCH) model to capture the leverage effect.}

Let $\rho_{ij|\mathbf{z}}^{[t]}=\text{Corr}(X_{it},X_{jt}|\{ X_{kt}: k\in \mathbf{z} \},\mathcal{F}_{t-1})$, where $\mathcal{F}_{t-1}$ is the information set up to time $t-1$. Each element in the set of conditional correlations at time $t$, $\{\rho_{i,i_t[k]|i_t[k-1],\ldots,i_t[1]}^{[t]}:k=1,\ldots,n_t(i)\land i=1,\ldots,n\}$, follows the DCC dynamics, where $i_t[k]$ is the $k$th parent of node $i$ in $G_t$ and $n_t(i)$ is the number of parents of node $i$ in $G_t$. Traditionally, the orders of the parents in the conditional correlations are decided according to a topological order $\prec_t$ of $G_t$. As there could be multiple possible topological orders of $G_t$, whenever there is a subset of nodes that are exchangeable in a topological order, we order the nodes according to their volatilities in \eqref{eqt:GARCH}. The rationale behind this is that we assume that a more volatile stock should be more influential. For example, suppose that the network in \autoref{fig:network_eg_step1} is our network at time $t$. Then, $(1,2,3,4,5)$ and $(2,1,3,4,5)$ are two possible topological orders. Nodes 1 and 2 are exchangeable. Suppose that $\sigma_{1t}^2 > \sigma_{2t}^2$, then, we handle node 1 first and the set of conditional correlations is chosen to be $\{ \rho_{3,1}^{[t]}, \rho_{3,2|1}^{[t]}, \rho_{4,3}^{[t]}, \rho_{5,4}^{[t]} \}$ (then, $3_t[1]=1$ and $3_t[2]=2$). Similarly, if $\sigma_{1t}^2 < \sigma_{2t}^2$, we choose $\{ \rho_{3,2}^{[t]}, \rho_{3,1|2}^{[t]}, \rho_{4,3}^{[t]}, \rho_{5,4}^{[t]} \}$ (then, $3_t[1]=2$ and $3_t[2]=1$). Specifically, the DCC dynamics is given by
\begin{equation}
	\begin{aligned}
		\rho_{i,i_t[k]|i_t[k-1],\ldots,i_t[1]}^{[t]} =& (1-a^{C}-b^{C})\overline{\rho}_{i,i_t[k]|i_t[k-1],\ldots,i_t[1]} + a^C \xi_{i,i_t[k]|i_t[k-1],\ldots,i_t[1]}^{[t-1]} +\\ & b^C\rho_{i,i_t[k]|i_t[k-1],\ldots,i_t[1]}^{[t-1]},
		\label{eqt:cond_corr}
	\end{aligned}
\end{equation}
where $\overline{\rho}_{i,i_t[k]|i_t[k-1],\ldots,i_t[1]}$ is the long-run conditional correlation,  $\xi_{i,i_t[k]|i_t[k-1],\ldots,i_t[1]}^{[t-1]}$ is a sample conditional correlation, calculated by
$$
\xi_{ij|\mathbf{z}}^{[t-1]}  = \frac{\sum_{\tau=1}^{m_c} x_{i|\mathbf{z},t-\tau}x_{j|\mathbf{z},t-\tau} }{\sqrt{\sum_{\tau=1}^{m_c}  x_{i|\mathbf{z},t-\tau}^2\sum_{\tau=1}^{m_c}  x_{j|\mathbf{z},t-\tau}^2}},
$$
with $x_{i|\mathbf{z},t-\tau}=X_{i,t-\tau}-\bo\beta_{i|\mathbf{z},t-\tau}^T \mathbf{Z}_{t-\tau}$, $\bo\beta_{i|\textbf{z},t-\tau}=\left(\Sigma_{\mathbf{Z}\mathbf{Z},t-\tau}\right)^{-1}\bo\sigma_{\mathbf{Z}i,t-\tau}$ for any $i\in\{1,\ldots,n\}$, vector $\mathbf{z}\subseteq \{1,\ldots,n\}\setminus\{i,j\}$, and $\mathbf{Z}_{t-\tau}$ is a column vector containing the elements in the set $\{X_{k,t-\tau}: k\in \mathbf{z}\}$. $\Sigma_{\mathbf{Z}\mathbf{Z},t-\tau}$ is the covariance matrix of the nodes in $\mathbf{Z}_{t-\tau}$, and $\boldsymbol{\sigma}_{\mathbf{Z}i,t-\tau}$ is the column vector containing the covariances between $X_{i,t-\tau}$ and nodes in $\mathbf{Z}_{t-\tau}$. Equivalently, $x_{i|\mathbf{z},t-\tau}$ is the residual in the regression of $X_{i,t-\tau}$ on $\mathbf{Z}_{t-\tau}$.   We take $m_c=2$ and initialize $\xi_{ij|\mathbf{z}}^{[1]}=\overline{\rho}_{ij|\mathbf{z}}$. When the conditional correlations $\{\rho_{i,i_t[k]|i_t[k-1],\ldots,i_t[1]}^{[t]}:k=1,\ldots,n_t(i)\land i=1,\ldots,n\}$ are computed from \eqref{eqt:cond_corr}, the unconditional correlation matrix $R_t$ can be calculated using the recursion formula \citep{vine_copula}

$$
\rho_{ij|\mathbf{z}_{-k}}^{[t]} = \rho_{ij|\mathbf{z}}^{[t]} \sqrt{[1-(\rho_{i{z}_{k}|\mathbf{z}_{-k}}^{[t]})^2][1-(\rho_{j{z}_{k}|\mathbf{z}_{-k}}^{[t]})^2]} + \rho_{i{z}_{k}|\mathbf{z}_{-k}}^{[t]}\rho_{j{z}_{k}|\mathbf{z}_{-k}}^{[t]},
$$
where $\mathbf{z}_{-k}$ is the vector $\mathbf{z}$ without the $k$th element, and ${z}_k$ is the $k$th element of $\mathbf{z}$. The term of the form $\overline{\rho}_{xy|\mathbf{z}}$ in the intercept in \eqref{eqt:cond_corr} is obtained from the constant correlation matrix $\overline R=[\overline{\rho}_{ij}]_{i,j=1}^n$ by computing the precision matrix of the subset of random variables corresponding to the columns $x$, $y$, and $\mathbf{z}$ in $\overline{R}$ \citep{lauritzen1996graphical}. 


\subsection{Exogenous Variables for Conditional Addition and Deletion Dynamics and Edge Selection Dynamics}
We use a market volatility index to drive the dynamics of $a_t$ and $d_t$ in \eqref{eqt:dynamic_mean_spec}. Specifically, we include the centered price of the volatility index at time $t-1$, i.e., $g_a(v_{t-1})=\gamma_1 (v_{t-1}-\overline{v})$ and $g_d(v_{t-1})=\gamma_2 (v_{t-1}-\overline v)$, where $\overline v = \sum_{t=1}^T v_{t}/T$, $v_{t}$ is the price of the volatility index on day $t$ and $\gamma_1$ and $\gamma_2$ are two parameters. For the activeness dynamics in \eqref{eqt:dynamics_wit_reference_group}, we choose $x_{i,t-1}$ to be the volatility of the $i$th stock relative to the $n$th stock, i.e., we set $g_{ES}(\sigma_{i,t}^2)= \sigma_{i,t}^2-\sigma_{n,t}^2 $, as the model only compares the magnitudes among $w_{i,t}$'s, we need an identifiability constraint so that $w_{i,t}$'s are estimable. We set the $n$th stock as the reference group so that $w_{n,t}=0.5$ on any day $t$. Then, $w_{i,t}>0.5$ ($w_{i,t}<0.5$) implies that the $i$th stock is more (less) active than the $n$th stock. These choices are based on the empirical evidence that volatile markets are often accompanied by unstable network changes \citep{Lupe2023}. {Another choice is to use the liquidity of a stock to drive the dynamics of $w_{i,t}$'s, since a stock with high liquidity can be sold more rapidly and may have a higher influence on the stock market. }

%


\subsection{Posterior Distributions}
The parameters in the DBN-AD model are (1) the initial network: $G_1$, (2) parameters in the edge change dynamics: $\alpha_1$, $\alpha_2$, $\beta_1$, $\beta_2$, $\gamma_1$, $\gamma_2$, $\overline{\mu}^a$ and $\overline{\mu}^d$, (3) the numbers of edge addition and deletion: $\mathbf{a}=\{a_t\}_{t=2}^T$ and $\mathbf{d}=\{d_t\}_{t=2}^T$, (4) parameters in the edge selection: Initial weights $\{w_{i,1}\}_{i=1}^n$ and $\beta^{ES}$, (5) parameters in the GARCH processes: $\{\alpha_{i}^G\}_{i=1}^n$, $\{\beta_{i}^G\}_{i=1}^n$ and $\{\overline{\sigma}_i^2\}_{i=1}^n$, and (6) Parameters in the conditional correlation dynamics: $\overline R$, $a^C$ and $b^C$. We denote the set containing all these parameters as $\Theta$. Let $\mathcal{D}_t$ be the data on day $t$, $\mathcal{D}=\bigcup_{t=1}^T \mathcal{D}_t$ and $\mathcal{F}=\bigcup_{t=1}^{T-1} \mathcal{F}_t$. The joint posterior distribution of these parameters is
{
	\scriptsize
	\begin{align*}
		&P(\overline R,a^C,b^C, G_1 ,\mathbf{a},\mathbf{d},\overline{\mu}^a,\overline{\mu}^d,\alpha_1,\beta_1,\alpha_2,\beta_2,\gamma_1,\gamma_2,\{w_{i,1}\}_{i=1}^n,\beta^{ES},\{\alpha_{i}^G\}_{i=1}^n, \{\beta_{i}^G\}_{i=1}^n, \{\overline{\sigma}_i^2\}_{i=1}^n|\mathcal{D},\mathcal{F})\\
		\propto& P(\mathcal{D}|\Sigma_1,\ldots,\Sigma_T,\mathcal{F})\cdot P(\mathbf{a},\mathbf{d}|\overline{\mu}^a,\overline{\mu}^d,\alpha_1,\beta_1,\alpha_2,\beta_2,\gamma_1,\gamma_2,\mathcal{F})\cdot \\ & P(\overline R,a^{C},b^{C}, G_1 ,\overline{\mu}^a,\overline{\mu}^d,\alpha_1,\beta_1,\alpha_2,\beta_2,\gamma_1,\gamma_2,\{w_{i,1}\}_{i=1}^n,\beta^{ES},\{\alpha_{i}^G\}_{i=1}^n, \{\beta_{i}^G\}_{i=1}^n, \{\overline{\sigma}_i^2\}_{i=1}^n) \\
		= &P(\mathcal{D}_1|\Sigma_1) \prod_{t=2}^T\left[ P(\mathcal{D}_t|\Sigma_t,\mathcal{F}_{t-1})\cdot P(a_t|\mu_t^a,a_{max,t})  P(d_t|\mu_t^d,d_{max,t})\right] P(\overline R)P(a^{C},b^{C})P( G_1) P(\overline{\mu}^a,\overline{\mu}^d)\cdot \\ &  P(\alpha_1,\beta_1,\alpha_2,\beta_2,\gamma_1,\gamma_2) P(\{w_{i,1}\}_{i=1}^n,\beta^{ES}) P(\{\alpha_{i}^G\}_{i=1}^n, \{\beta_{i}^G\}_{i=1}^n, \{\overline{\sigma}_i^2\}_{i=1}^n). \numberthis \label{eqt:new_scheme_poseterior_2}
	\end{align*}
}
As the possible numbers of edges to be added and deleted are finite and dependent on the network $G_t$, we denote $a_{max,t}$ and $d_{max,t}$ be respectively the maximum values, and set $a_t=a_{max,t}$ whenever the sampled value from $\text{Poisson}(\mu_t^a)$ is greater than $a_{max,t}$, and similarly for $d_t$. The corresponding probabilities are given by

$$
P(a_t|\mu_t^a,a_{max,t}) =  \left( \frac{e^{-\mu_t^a}(\mu_t^a)^{a_t}}{a_t!} \right)^{\mathbbm{1}(a_t  <  a_{max,t})}
\left( \sum_{x=a_{max,t}}^{\infty} \frac{e^{-\mu_t^a}(\mu_t^a)^{x}}{x!}
\right)^{\mathbbm{1}(a_t  =  a_{max,t})},
$$
and 
$$
P(d_t|\mu_t^d,d_{max,t}) =  \left( \frac{e^{-\mu_t^d}(\mu_t^d)^{d_t}}{d_t!} \right)^{\mathbbm{1}(d_t  <  d_{max,t})}
\left( \sum_{x=d_{max,t}}^{\infty} \frac{e^{-\mu_t^d}(\mu_t^d)^{x}}{x!}
\right)^{\mathbbm{1}(d_t  =  d_{max,t})}.
$$

Assuming the parameters in different equations are independent a priori, the priors are as follows:
\begin{enumerate}[(i)]
	\item $P(\overline R)\propto 1$ if $\overline R$ is a positive definite matrix with all ones in the diagonals, and zero otherwise.
	\item $P(a^{C},b^{C})\propto 1$ if $a^{C},b^{C}\geq 0$ and $a^{C}+b^{C}<1$, and zero otherwise.
	\item $P( G_1)\propto \mathbbm{1}(G_1\text{ is a DAG})$, where $\mathbbm{1}(A)=1$ if the event $A$ holds, and zero otherwise.
	\item $P(\overline{\mu}^a,\overline{\mu}^d)\propto 1$ if $\overline{\mu}^a$ and $\overline{\mu}^d$ are positive.
	\item $P(\alpha_1,\beta_1,\alpha_2,\beta_2,\gamma_1,\gamma_2)\propto 1$ if $\alpha_j,\beta_j\geq 0$, $\alpha_j+\beta_j<1$ for $j=1,2$, and zero otherwise. There is no restriction imposed on $\gamma_1$ and $\gamma_2$.
	\item $P(\{w_{i,1}\}_{i=1}^n,\beta^{ES}) \propto 1$ if $0\leq \beta^{ES}\leq 1$ and $w_{i,1}\in (0,1)$ for all $i=1,\ldots,n$.
	\item  $P(\{\alpha_{i}^G\}_{i=1}^n, \{\beta_{i}^G\}_{i=1}^n, \{\overline{\sigma}_i^2\}_{i=1}^n)\propto 1$ if $\alpha_i^G,\beta_i^G\geq 0$, $\alpha_i^G+\beta_i^G<1$ and $\overline{\sigma}_i^2>0$ for all $i=1,\ldots,n$, and zero otherwise.
\end{enumerate}
Note that, in the MCMC sampling, we further restrict that $\beta_1,\beta_2>0.5$. Since we only have one observation per time point, the posterior distribution could be highly multimodal in networks, and thus there could be many networks that fit the data equally well. Without imposing such restriction, the sampler could search for a time series of networks with no clear dynamics (low transition probabilities) but fitting the data well (high likelihood scores), leading to the overfitting problem. Setting $\beta_1,\beta_2>0.5$ can promote a more efficient search for the network dynamics.

\subsection{Network Initialization}
\label{section:network_initialization}
The MCMC sampling is started with the time series of networks obtained by a moving-window method using the Bayesian Gaussian equivalent (BGe) score \citep{Heckerman1995LearningBN} as the score function. The window size is 30 time points. To ensure that the time series of networks is smooth, we add a penalty prior
\begin{equation}
	p(G_t|G_{t-1}) = \exp(-\lambda\norm{G_t-G_{t-1}}),
	\label{eqt:ad_hoc_prior}
\end{equation}
where
$$
\norm{A} = \sum_{i=1}^n \sum_{j=1}^n |a_{ij}|,
$$
for any $n$ by $n$ matrix $A$ whose $(i,j)$th element is $a_{ij}$ and for some $\lambda>0$. Note that the prior in \eqref{eqt:ad_hoc_prior} is for network initialization only. $\lambda$ cannot be too small, otherwise, the number of edge changes between consecutive networks will be very large; if $\lambda$ is too large, the penalty is too strong and thus there will be only a few or even no edge changes between consecutive networks. Our experiments suggest that $\lambda \in [n/10,n/5]$ performs well.  To improve the mixing of the MCMC, we adopt the multi-step proposal for MCMC structure learning \citep{larjo2015using}. The proposal consists of $M$ steps of single-edge movements, where $M$ is sampled from $\{1,2,3,4,5\}$ randomly for each iteration. We set the maximum of $M$ as 5 as the network is harder to accept if the number of edge moves is too large. 

\subsection{Parameter Estimation and Structural Learning}

\subsubsection{Sampling Scheme for $G_1$, $\{a_t\}_{t=2}^T$ and $\{d_t\}_{t=2}^T$}
\label{section:sampling_for_at_dt}
We again use the multi-step proposal for $G_1$ as described in Section \ref{section:network_initialization}. The subsequent networks $\{G_2,\ldots,G_T\}$ are not sampled directly; instead, we conduct sampling on the numbers of edge addition and deletion $\{a_t\}_{t=2}^T$ and $\{d_t\}_{t=2}^T$. To enhance the mixing, we conduct block sampling for $\{a_t\}_{t=2}^T$ and $\{d_t\}_{t=2}^T$ and we use two different proposals. The block size is set to at most 10, as a larger block size leads to a poor acceptance rate. The first proposal uses a random-walk move for a local change, and the second proposal uses a cyclic move for a global change.

\begin{enumerate}[(1)]
	\item (Random-walk move) A random block size of $B$ is drawn from $\{1,\ldots,10\}$ and a time point $\tau$ is drawn from $\{2,\ldots,T-B+1\}$ (note that we do not have $a_1$ and $d_1$). We sample $a_t$ and $d_t$ for the $B$ time points $t=\tau,\tau+1,\ldots,\tau+B-1$. Let $a_t^{(m)}$ be the $m$th sample of $a_t$ in the MCMC sampling. If $a_t^{(m)}=0$, $a_t'$ is sampled from $\{0,1\}$ randomly; if $a_t^{(m)}>0$, we set $a_t'=a_t^{(m)}+U$, where $U$ is sampled from $\{-1,0,1\}$ randomly. The ratio of transition probabilities is
	$$
	\frac{p(a_t^{(m)}|a_t')}{p(a_t'|a_t^{(m)})}=
	\begin{cases}
		\frac{1/2}{1/2}=1 & \text{ if $a_t^{(m)}=0$ and $a_t'=0$,}\\
		\frac{1/2}{1/3}=3/2 & \text{ if $a_t^{(m)}=1$ and $a_t'=0$,}\\
		\frac{1/3}{1/2}=2/3 & \text{ if $a_t^{(m)}=0$ and $a_t'=1$,}\\
		\frac{1/3}{1/3}=1 & \text{ if $a_t^{(m)},a_t' \in \mathbb{Z}^+$ and $(a_t^{(m)}-a_t')\in\{-1,0,1\}$,}\\
		0 & \text{ otherwise.}
	\end{cases}
	$$
	The sampling scheme for $d_t$ is similar.
	
	\item (Cyclic move) A random block size of $B$ is drawn from $\{2,\ldots,10\}$ and a time point $\tau$ is drawn from $\{2,\ldots,T-B+1\}$. For $a_t$ and $d_t$, $t=\tau,\tau+1,\ldots,\tau+B-1$, we have a probability of 1/2 to shift all elements to the right, i.e., we set $(a_\tau',a_{\tau+1}',\ldots,a_{\tau+B-1}')$ to $(a_{\tau+B-1}^{(m)},a_{\tau}^{(m)},a_{\tau+1}^{(m)},\ldots,a_{\tau+B-2}^{(m)})$ and $(d_\tau',d_{\tau+1}',\ldots,d_{\tau+B-1}')$ to $(d_{\tau+B-1}^{(m)},d_{\tau}^{(m)},d_{\tau+1}^{(m)},\ldots,$ $d_{\tau+B-2}^{(m)})$, where the last elements are relocated at the beginning of the sets. With a probability of 1/2, we shift all elements to the left, i.e., we set $(a_\tau',a_{\tau+1}',\ldots,a_{\tau+B-1}')$ to $(a_{\tau+1}^{(m)},a_{\tau+2}^{(m)},\ldots,a_{\tau+B-1}^{(m)},a_{\tau}^{(m)})$ and $(d_\tau',d_{\tau+1}',\ldots,d_{\tau+B-1}')$ to $(d_{\tau+1}^{(m)},d_{\tau+2}^{(m)},\ldots,$ $d_{\tau+B-1}^{(m)},$ $d_{\tau}^{(m)})$.  The ratios of transition probabilities are evaluated as follows (we omit the superscript ${(m)}$ in the following part to save notation):
	\begin{enumerate}[(i)]
		\item If $B$ is even, and $a_t$ and $d_t$ for $t=\tau,\tau+1,\ldots,\tau+B-1$ do not repeat themselves by shifting left by two steps and right by two steps, i.e., $a_\tau=a_{\tau+2}=\ldots=a_{\tau+B-2}$,  $a_{\tau+1}=a_{\tau+3}=\ldots=a_{\tau+B-1}$,  $d_\tau=d_{\tau+2}=\ldots=d_{\tau+B-2}$ and  $d_{\tau+1}=d_{\tau+3}=\ldots=d_{\tau+B-1}$ do not hold. Then, if we shift all elements to left, we need to shift them to the right to get back to the original $\mathbf{a}$ and $\mathbf{d}$. Then, the ratio of transition probabilities is $(1/2)/(1/2)=1$. 
		\item  If $B$ is even and $a_\tau=a_{\tau+2}=\ldots=a_{\tau+B-2}$,  $a_{\tau+1}=a_{\tau+3}=\ldots=a_{\tau+B-1}$,  $d_\tau=d_{\tau+2}=\ldots=d_{\tau+B-2}$ and  $d_{\tau+1}=d_{\tau+3}=\ldots=d_{\tau+B-1}$ hold, then, the ratio of transition probabilities is $1/1=1$ as the sequence $\{a_t,d_t\},\{a_{t+1},d_{t+1}\},\ldots,$ $\{a_{t+B-1},$ $d_{t+B-1}\}$ repeats themselves with any shift of 2 steps.
		\item If $B$ is odd and not all $\{a_t,d_t\},\{a_{t+1},d_{t+1}\},\ldots,\{a_{t+B-1},d_{t+B-1}\}$ are identical, the ratio of transition probabilities is $(1/2)/(1/2)=1$ as we need to reverse the shift to get back the original $\mathbf{a}$ and $\mathbf{d}$.
		\item  If $\{a_t,d_t\},\{a_{t+1},d_{t+1}\},\ldots,\{a_{t+B-1},d_{t+B-1}\}$ are identical, then the move makes no change in $\mathbf{a}$ and $\mathbf{d}$.
	\end{enumerate}
\end{enumerate}

\subsubsection{Sampling of $\overline{R}$: Parameter-Extended Metropolis-Hastings (PX-MH)}
To sample the correlation matrix $\overline{R}$, we adapt the Parameter-extended Metropolis-Hastings (PX-MH) algorithms \citep{zhang2006sampling}. First note that the diagonals of the matrix sampled from the Wishart distribution may not be 1. Then, we cannot directly sample correlation matrices from the Wishart distribution; instead, we sample the covariance matrix $\Sigma = D^{\frac{1}{2}}RD^{\frac{1}{2}}$ from the Wishart distribution with scale parameter $\mathbf{V}$ and degrees of freedom $N$. The joint density function of $(R,D)$ is
$$
f_{R,D}(R,D;N,\mathbf{V})= f_\Sigma(D^{\frac{1}{2}}RD^{\frac{1}{2}};N,\mathbf{V}) \cdot J,
$$
where $f_\Sigma(\Sigma;N,\mathbf{V})$ is the density function of $\text{Wishart}_n(N,\mathbf{V})$, i.e., the Wishart distribution with degrees of freedom $N>n$ and a $n\times n$ positive definite scale matrix $\mathbf{V}$, given by
$$
f_\Sigma(\Sigma;N,\mathbf{V}) = \frac{| \Sigma |^{(N-n-1)/2} \exp\left( -\frac{1}{2}\text{tr}(\mathbf{V}^{-1}\Sigma)  \right)}{2^{\frac{Nn}{2}} 
	|\mathbf{V}|^{\frac{N}{2}}\Gamma_n\left(\frac{N}{2}\right)
},
$$
where $\Gamma_n$ is the multivariate gamma function, and  $J$ is the Jacobian of the transformation $\Sigma\to (R,D)$, given by
$$
J = \prod_{i=1}^n d_i^{\frac{n-1}{2}},
$$
$d_i$ is the $i$th diagonal element in $D$.\\

The steps for the MCMC sampling of the constant correlation matrix $\overline R$ are as follows. To save notation, we write $R$ instead of $\overline R$ in the following algorithm. We first set an initial covariance matrix $\Sigma^{(0)}=[D^{(0)}]^{\frac{1}{2}} R^{(0)}[D^{(0)}]^{\frac{1}{2}}$.  The initial sample is $( R^{(0)},D^{(0)})$. In each iteration $m$:
\begin{enumerate}
	\item Generate $( R',D')$ by generating $\Sigma'=D'^{\frac{1}{2}} R'D'^{\frac{1}{2}}$ from $\text{Wishart}_n(N,\Sigma^{(m)}/N)$.
	\item Set $( R^{(m+1)},D^{(m+1)})=( R',D')$ with a probability of
	$$
	\min\left\{  \frac{
		P( R'|\mathcal{D},\mathcal{F},\Theta^{(m)}\setminus\{ R^{(m)}\})
	}{
		P( R^{(m)}|\mathcal{D},\mathcal{F},\Theta^{(m)}\setminus\{ R^{(m)}\})
	} \cdot \frac{
		p( R^{(m)},D^{(m)}| R',D')
	}{
		p( R',D'| R^{(m)},D^{(m)})  
	},1 \right\},
	$$
	and set $( R^{(m+1)},D^{(m+1)})=( R^{(m)},D^{(m)})$ otherwise. $\Theta^{(m)}$ is the set of all parameters in iteration $m$ and $p( R',D'| R^{(m)},D^{(m)})$ is the transition kernel, given by
	$$
	p( R',D'| R^{(m)},D^{(m)}) = f_{R,D}\left( R',D';N, \frac{ [D^{(m)}]^{\frac{1}{2}} R^{(m)} [D^{(m)}]^{\frac{1}{2}} }{N}\right).
	$$
	The full conditional distribution $P( R'|\mathcal{D},\mathcal{F},\Theta^{(m)}\setminus\{ R^{(m)}\})$ is obtained from the posterior distribution in \eqref{eqt:new_scheme_poseterior_2}:
	$$
	\begin{aligned}
		P( R'|\mathcal{D},\mathcal{F},\Theta^{(m)}\setminus\{ R^{(m)}\}) \propto \left[\prod_{t=1}^T P(\mathcal{D}_t|\Sigma_t')\right] P({R}'),
	\end{aligned}
	$$ 
	where $\Sigma_t'$ is the covariance matrix at time $t$, evaluated from the proposal correlation matrix ${R}'$ and using the methodology in Section \ref{section:volatility_and_correlation}. Note that we set the scale matrix $\mathbf{V}$ as $\Sigma^{(m)}/N$ so that $E(\Sigma')=N\cdot \Sigma^{(m)}/N=\Sigma^{(m)}$. Since $Var(\Sigma')\propto N^{-1}$, we adaptively tune $N$ in the MCMC sampling to achieve the optimal acceptance rate.

\end{enumerate}

\subsubsection{Sampling Scheme for the Remaining Parameters}
\label{section:RAM}
For the remaining parameters, we block them into groups and conduct the robust adaptive metropolis (RAM) \citep{RAM}. We let $\theta$ be a vector of parameters of interest and $\theta^{(m)}$ be the $m$th sample of $\theta$. Starting at $m=1$:

\begin{enumerate}
	\item We set a proposal
	$$
	\theta' = \theta^{(m)}+S_{m}U_m,
	$$
	where $U_m$ is a $d$-dimension vector drawn from the proposal distribution, $d$ is the number of parameters in $\theta$, and $S_{m}$ is a matrix that captures the correlations among the parameters. We choose $U_m\sim N_d(\mathbf{0},I_d)$. 
	\item With a probability of $\alpha_m=\min\{P(\theta'| \Theta^{(m)}\setminus\{\theta^{(m)}\},\mathcal{D},\mathcal{F})/P(\theta^{(m)}|\Theta^{(m)}\setminus\{\theta^{(m)}\} ,$ $\mathcal{D},\mathcal{F}),1\}$, where $P(\cdot)$ is the posterior distribution, we set $\theta^{(m+1)}=\theta'$. Otherwise, we set $\theta^{(m+1)}=\theta^{(m)}$.
	\item Update $S_m$ by the lower-diagonal matrix with positive diagonal elements $S_{m+1}$ satisfying the equation
	$$
	S_{m+1}S_{m+1}^T = S_{m}\left(I+\eta_m(\alpha_m-\alpha_*)\frac{U_mU_m^T}{\norm{U_m}^2}\right)S_{m}^T,
	$$
	where $I$ is the $d$ by $d$ identity matrix, $\alpha_*\in(0,1)$ is the target mean acceptance probability of the algorithm, and $\{\eta_m\}_{m\geq 1}$ is a step size sequence decaying to zero, where $\eta_m\in (0,1]$. It can be done using Cholesky decomposition. We take $\alpha_*=0.234$ and $\eta_m = m^{-2/3}$ following \cite{RAM}.
	\item Set $m$ to $m+1$ and go back to step 1.
\end{enumerate}

We block the remaining parameters into the following groups for block samplings. We sample the transformed parameters (marked with overhead tildes if any) as shown below using RAM for better mixing in the MCMC sampling.
\begin{enumerate}[(1)]
	\item Parameters in the conditional correlation dynamics: $\{a^C,b^C\}$. $\widetilde{a}^C = \log[a^C/(1-a^C)]$ and $\widetilde{b}^C = \log[b^C/(1-b^C)]$.
	\item The parameters in the intercepts in the dynamics of $\mu_t^a$ and $\mu_t^d$: $\{\overline{\mu}^a,\overline{\mu}^d\}$. $\widetilde{\mu}^a = \log \overline{\mu}^a$ and $\widetilde{\mu}^d = \log \overline{\mu}^d$. 
	\item Slopes in the dynamics of $\mu_t^a$ and $\mu_t^d$: $\{\alpha_1,\beta_1,\alpha_2,\beta_2\}$. $\widetilde{\alpha}_j = \log[\alpha_j/(1-\alpha_j)]$ and $\widetilde{\beta}_j = \log[\beta_j/(1-\beta_j)]$ for $j=1,2$.
	\item Parameters related to the volatility in the dynamics of $\mu_t^a$ and $\mu_t^d$: $\{\gamma_1,\gamma_2\}$. No transformation is needed.
	\item Parameters in the node activeness: $\beta^{ES}$. $\widetilde{\beta}^{ES}=\log[\beta^{ES}/(1-\beta^{ES})]$.
	\item Slopes in the GARCH models: $\{\alpha_i^G,\beta_i^G\}_{i=1}^n$. $\widetilde{\alpha}_i^G = \log[\alpha_i^G/(1-\alpha_i^G)]$ and $\widetilde{\beta}_i^G = \log[\beta_i^G/(1-\beta_i^G)]$ for $i=1,\ldots,n$.
	\item Long-run variance in the GARCH model: $\{\overline{\sigma}_i^2\}_{i=1}^n$. $\widetilde{\overline{\sigma}}_i^2=\log \overline{\sigma}_i^2$ for $i=1,\ldots,n$.
	\item Initial node activeness: $\{w_{i,1}\}_{i=1}^n$. $\widetilde{w}_{i,1}=\log[w_{i,1}/(1-w_{i,1})]$ for $i=1,\ldots,n$.
\end{enumerate}

\section{Simulation Study}
\label{section:simulation_study}
We evaluate the estimation performance using 100 replications. We consider $n=20$ stocks and $T=250$ time points. The parameters regarding the network dynamics are the same for each replication, as shown in \autoref{fig:sim_par_est}. The initial networks, the number of edge changes $a_t$'s and $d_t$'s, the GARCH parameters, and the volatility indexes are, however, different in each simulation, to produce different realizations generated from the same source of network dynamics. The average running time is 2.6 days per replication.

\begin{figure}[H]
	\centering
	\includegraphics[width=13cm]{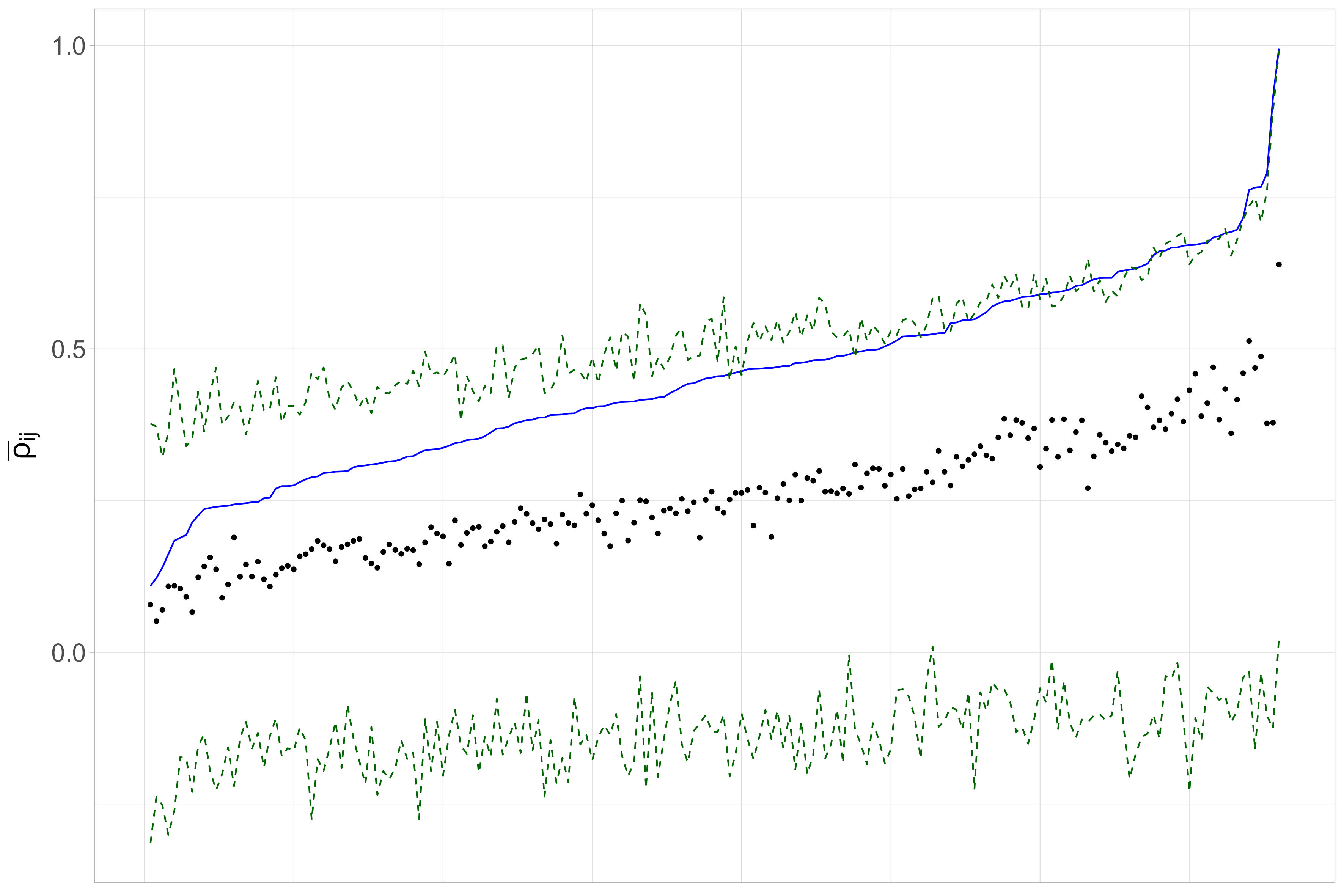}
	\caption{The medians of the posterior means of $\overline{\rho}_{ij}$'s. The blue line, black dots and the pair of green dotted lines represent respectively the true value, the medians, and the 95\% credible intervals of the medians.}
	\label{fig:source_correlation}
\end{figure}

The estimation results are presented in \autoref{fig:sim_par_est}. The median estimates out of 100 replications are close to their true values. The medians of the posterior mean estimates of $\overline{\rho}_{ij}$'s are plotted in \autoref{fig:source_correlation}. The estimates roughly follow the trend of the true values, while they are slightly underestimated. A reason is that, since the persistence $a^C+b^C$ is close to 1, $\overline{\rho}_{ij}$ has a small effect on the dynamic correlations are thus is more difficult to estimate. 

\begin{figure}[H]
	\centering
	\includegraphics[width=13cm]{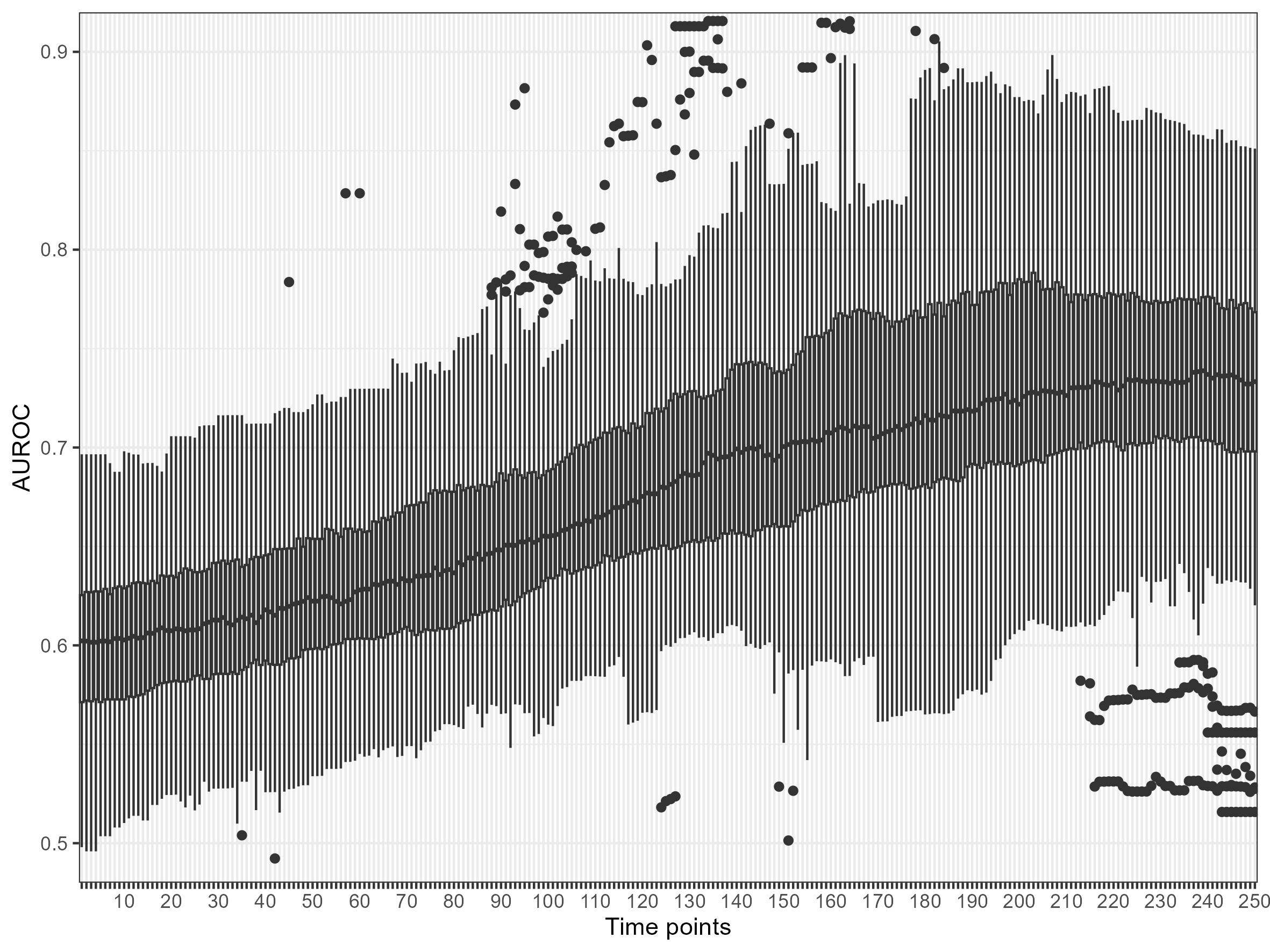}
	\caption{The boxplots showing the distributions of the AUROCs of the networks $G_t$ in the burn-in samples in all 100 replications, $t=1,\ldots,250$.}
	\label{fig:AUROC}
\end{figure}

\autoref{fig:AUROC} contains boxplots showing the distributions of the AUROCs of the networks $G_t$ in the burned-in samples in all 100 replications, $t=1,\ldots,250$. The networks are predicted relatively well near $t=250$, with an average AUROC of 0.744, whereas the performance near the starting points is relatively poor because the sampling of the networks near the left end time points is harder; A change in a network in that region will lead to changes in the networks in the subsequent time points, and thus a proposal is harder to accept. If time is available, we can enhance the estimation performance by running the MCMC sampling for longer.


\begin{table}[H]
	\centering
	\small
	\begin{tabular}{lllll}
		Parameter     & Median of posterior means & True value   & Lower 2.5\% & Upper 2.5\% \\ \hline
		$a^C$           & 0.0426           & 0.0603$~^*$  & 0.0268  & 0.0783 \\
		$b^C$           & 0.8857           & 0.877$~^*$   & 0.8121  & 0.9461 \\
		$\overline{\mu}^a$ & 0.0812           & 0.1$~^*$     & 0.0000  & 0.2503 \\
		$\overline{\mu}^d$ & 0.0727           & 0.1$~^*$     & 0.0002  & 0.2242 \\
		$\beta^{ES}$   & 0.3975           & 0.4055$~^*$  & 0.0341  & 0.7003 \\
		$\alpha_1$     & 0.0166           & 0.0215$~^*$  & 0.0000  & 0.2797 \\
		$\alpha_2$     & 0.0205           & 0.0215$~^*$  & 0.0000  & 0.2787 \\
		$\beta_1$      & 0.7795           & 0.9141$~^*$  & 0.5791  & 0.9950 \\
		$\beta_2$      & 0.8416           & 0.9141$~^*$  & 0.5595  & 0.9908 \\
		$\gamma_1$     & -0.0219          & 0.1266$~^*$  & -2.3864 & 0.4119 \\
		$\gamma_2$     & -0.1314          & -0.1372$~^*$ & -1.4652 & 0.2649
	\end{tabular}
	\caption{The median of the posterior means, the true values, and the 95\% credible intervals of the sampled parameters in the 100 replications. (*: the true value is covered by the 95\% credible interval)}
	\label{fig:sim_par_est}
\end{table}

\section{Empirical Study}

\label{section:empirical_fin_app_DBN}
The daily log-returns of the first 20 {from the 30 constituents} of the Dow Jones Industrial Average (DJIA) Index with the largest capitalizations are used in the empirical study. The {20} stocks consist of 91.9\% of the total capitalization of DJIA, and they are listed in \autoref{tab:stock_DBN}. Their log-returns from 5 March 2021 to 23 February 2024 (748 days of observations) are collected. We first fit the model using the first 250 days of observations, and the fitted model is used to predict the future 4 weeks' networks and covariance matrices. The model is updated every four weeks after the last trading day in the final week (i.e., we mimic the practice that models are updated at weekends) using the most recent $T=250$ days of return data. There are 25 updates throughout the period from 5 March 2021 to 23 February 2024. 

\subsection{A fitted model}
\begin{figure}[H]
	\centering
	\begin{subfigure}{0.95\textwidth}
		\centering
		\includegraphics[width=12cm]{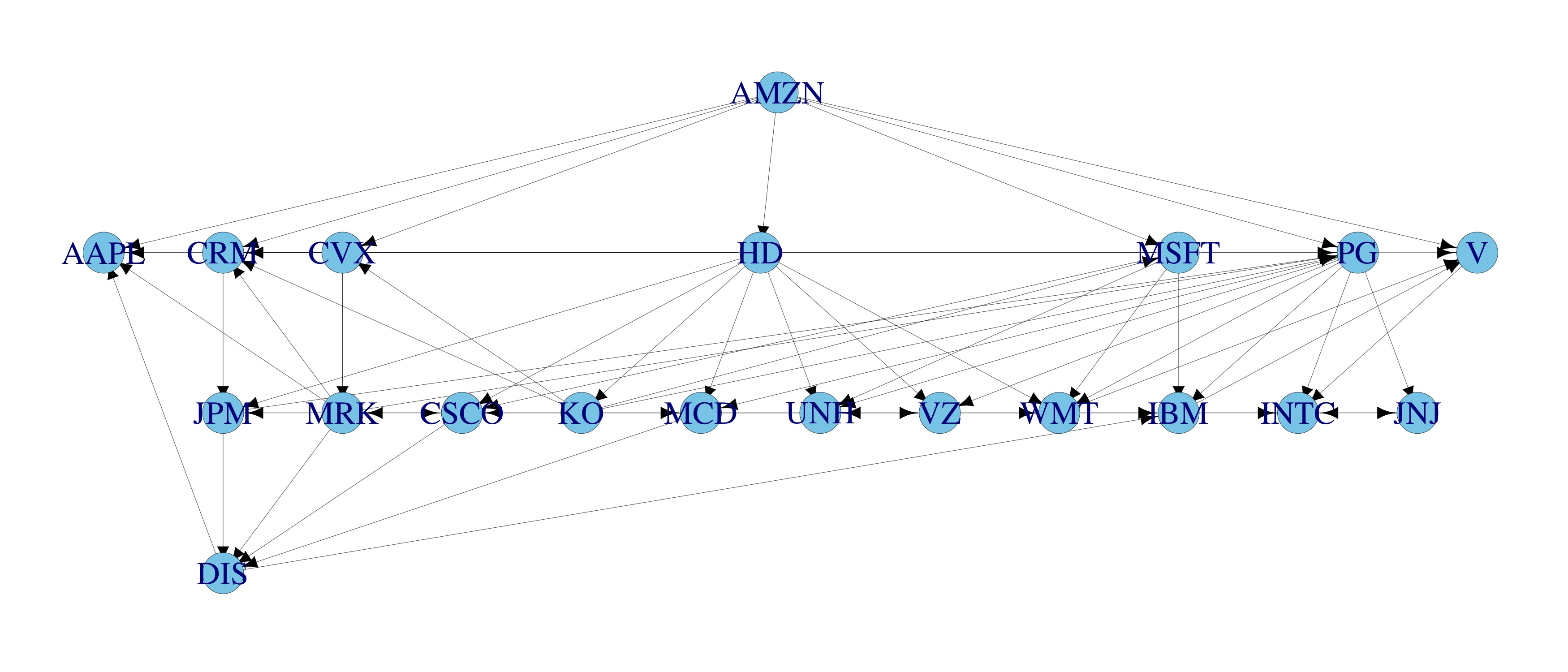}
		\caption{Network at time $t=1$ (March 5, 2021).}
	\end{subfigure}
	\begin{subfigure}{0.95\textwidth}
		\centering
		\includegraphics[width=12cm]{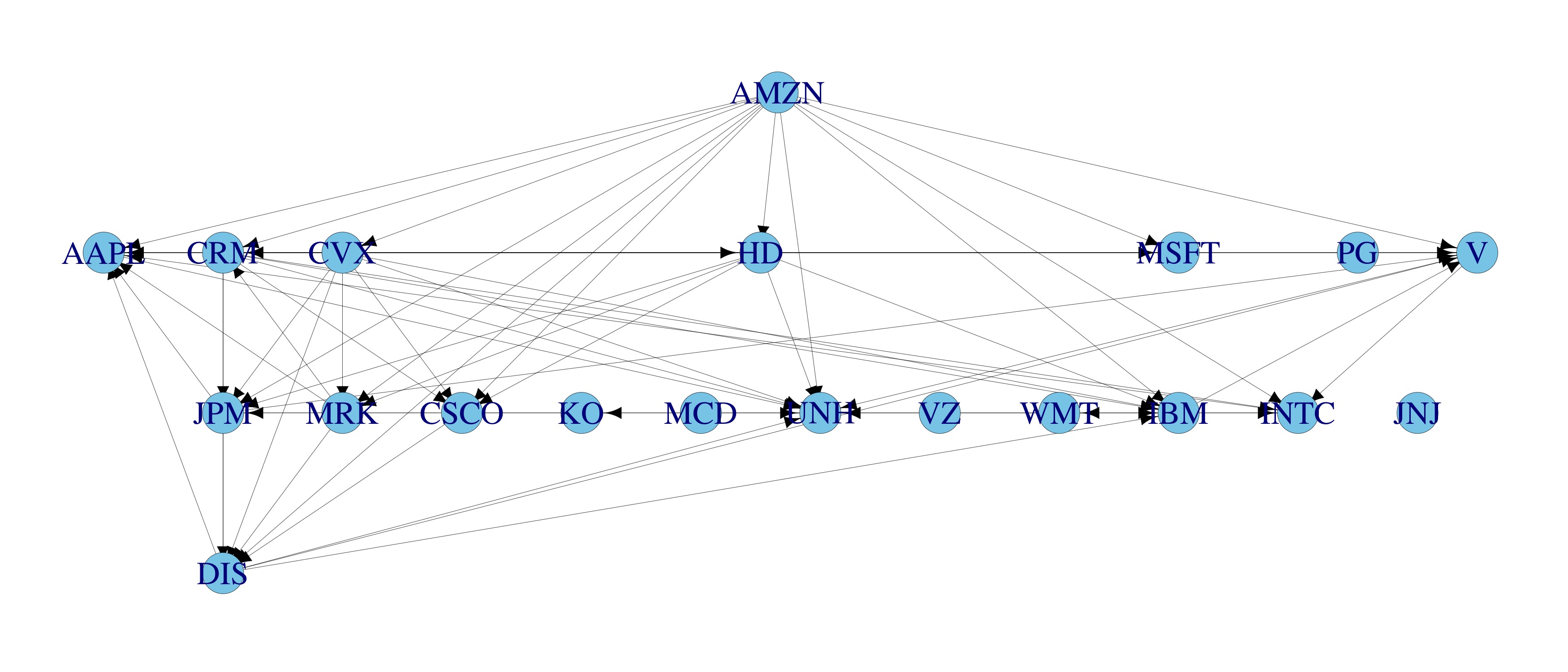}
		\caption{Network at time $t=125$ (August 31, 2021).}
	\end{subfigure}
	\begin{subfigure}{0.95\textwidth}
		\centering
		\includegraphics[width=12cm]{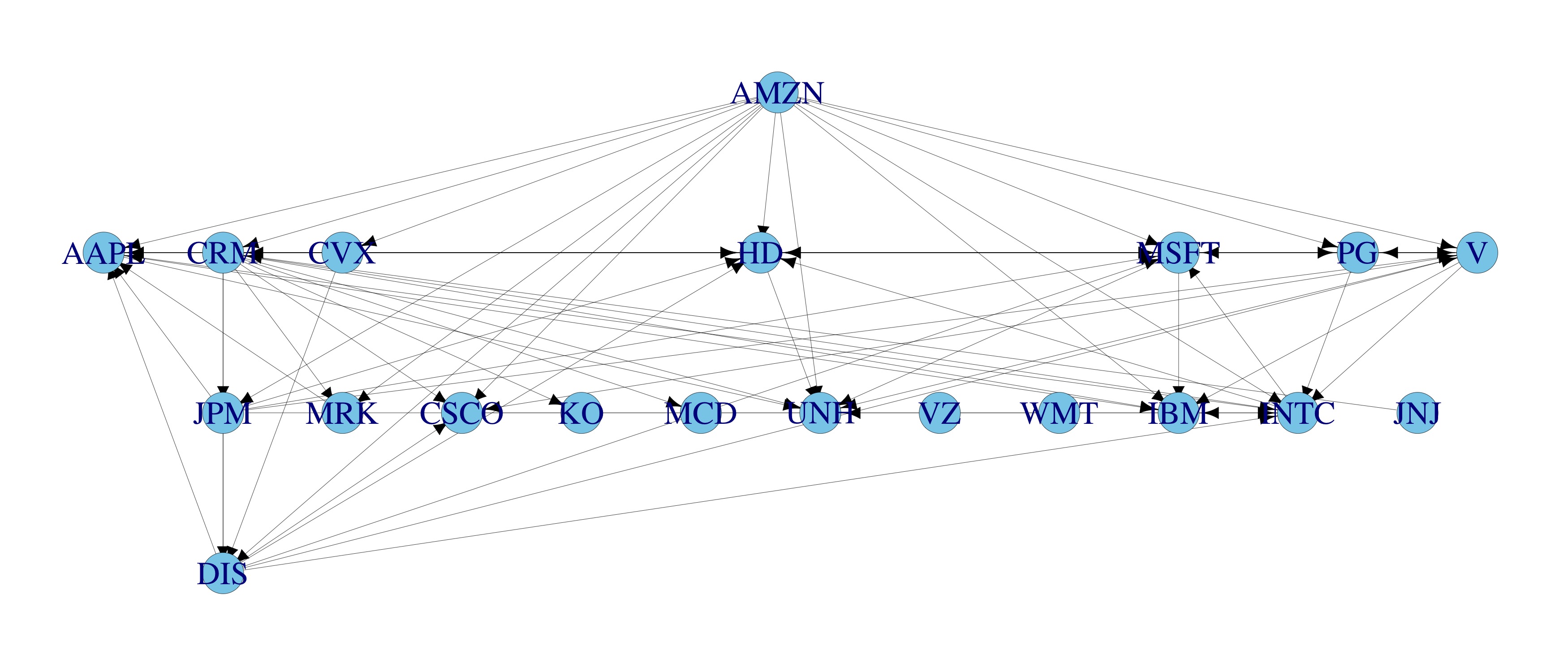}
		\caption{Network at time $t=250$ (March 1, 2022).}
	\end{subfigure}
	\caption{Three example networks learnt using the first 250 days of observations.}
	\label{fig:DBN_AD_example_networks}
\end{figure}

\begin{figure}[H]
	\centering
	\includegraphics[width=15cm]{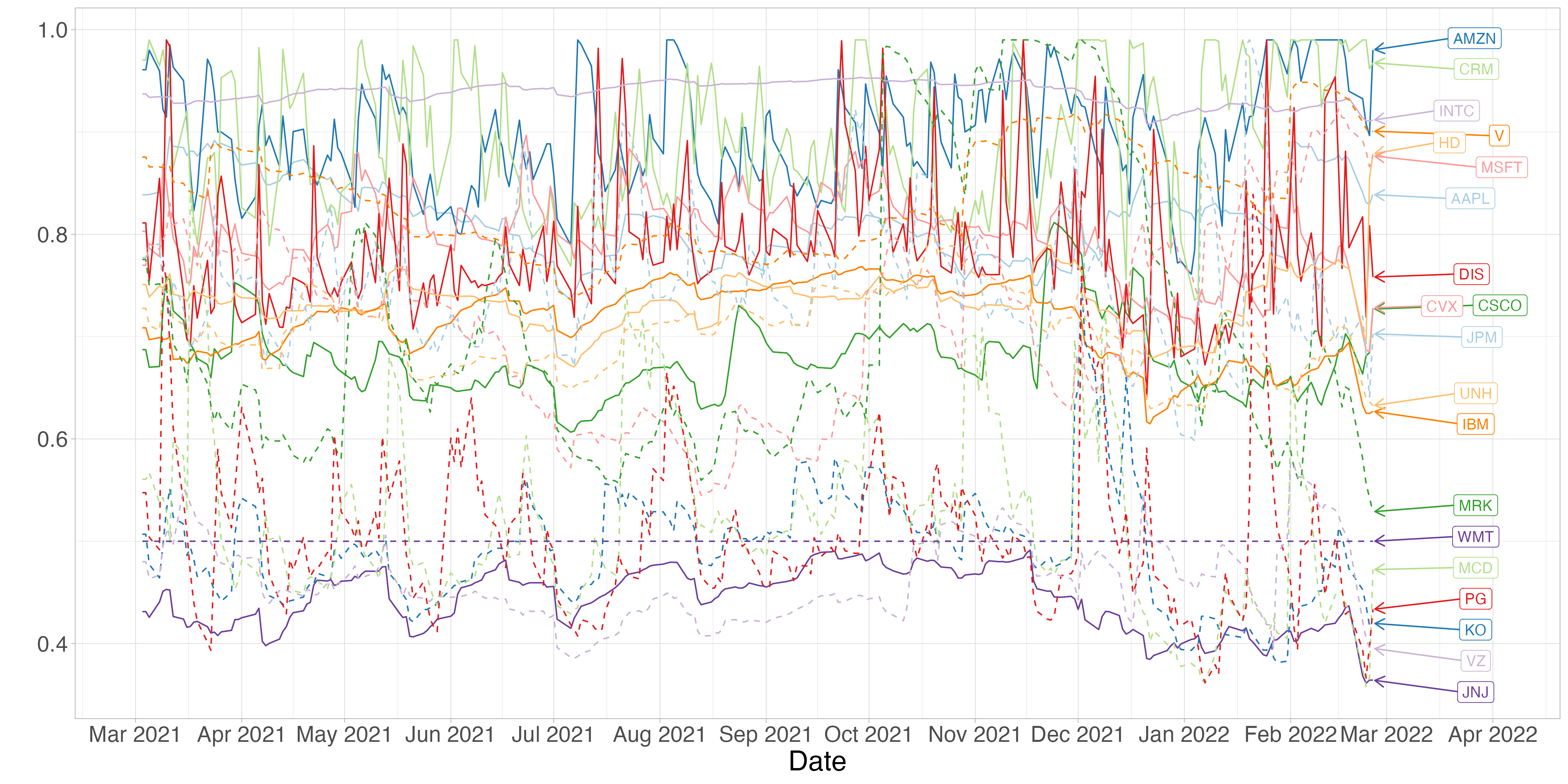}
	\caption{The time series of $w_{i,t}$ for $t=1,\ldots,250$ for each stock $i$, $i=1,\ldots,20$.}
	\label{fig:DBN_AD_wi}
\end{figure}

\begin{figure}[H]
	\centering
	\includegraphics[width=15cm]{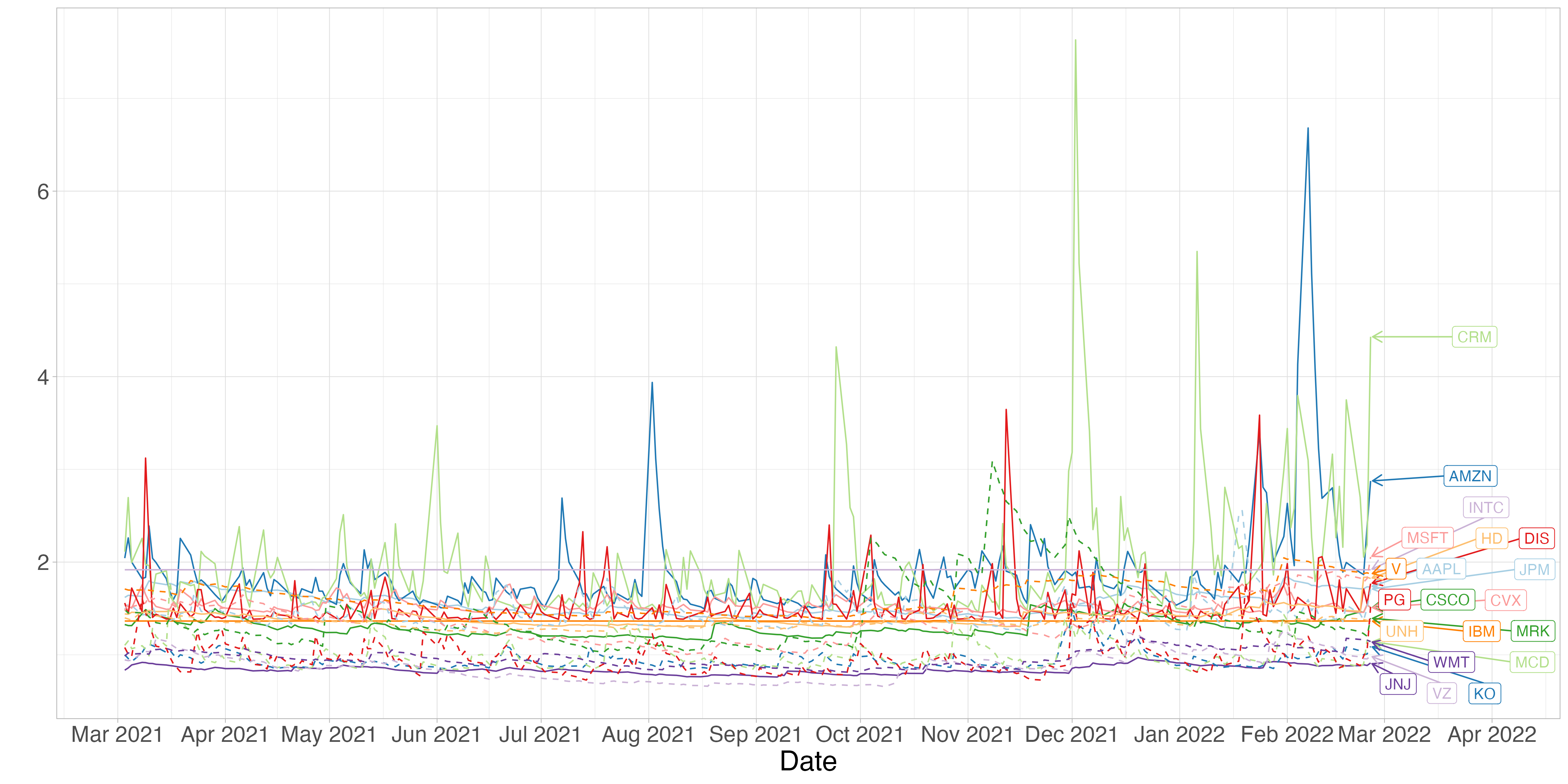}
	\caption{The time series of $\sigma_{i,t}$ for $t=1,\ldots,250$ for each stock $i$, $i=1,\ldots,20$.}
	\label{fig:DBN_AD_volatility}
\end{figure}

We first fit the DBN-AD model using the first 250 days of observation. We present the results of fitted networks and parameters to showcase the DBN-AD model. \autoref{fig:DBN_AD_example_networks} shows the learnt networks at time $t=1,125,250$, \autoref{fig:DBN_AD_wi} shows the time series of $w_{i,t}$, and \autoref{fig:DBN_AD_volatility} shows the times series of the volatility $\sigma_{i,t}$ for $t=1,\ldots,250$ for each stock $i$, $i=1,\ldots,20$. We observe that Amazon (AMZN) are ordered in top positions throughout the period of 250 days. More and more stocks are connected with AMZN over time. This is because AMZN has high levels of activeness over time, as observed in \autoref{fig:DBN_AD_wi}. Furthermore, we observe that all connections in AMZN are out-going; it is because the volatilities of AMZN are relatively large (as shown in \autoref{fig:DBN_AD_volatility}), and the model decides the directions of edges based on the volatilities. On the other hand, Johnson \& Johnson (JNJ) has only a few connections over time. It is due to the low levels of activeness and volatility shown respectively in \autoref{fig:DBN_AD_example_networks} and \autoref{fig:DBN_AD_volatility}. These two examples illustrate how connections change over time under DBN-AD model. 

\begin{figure}[H]
	\centering
	\includegraphics[width=12cm]{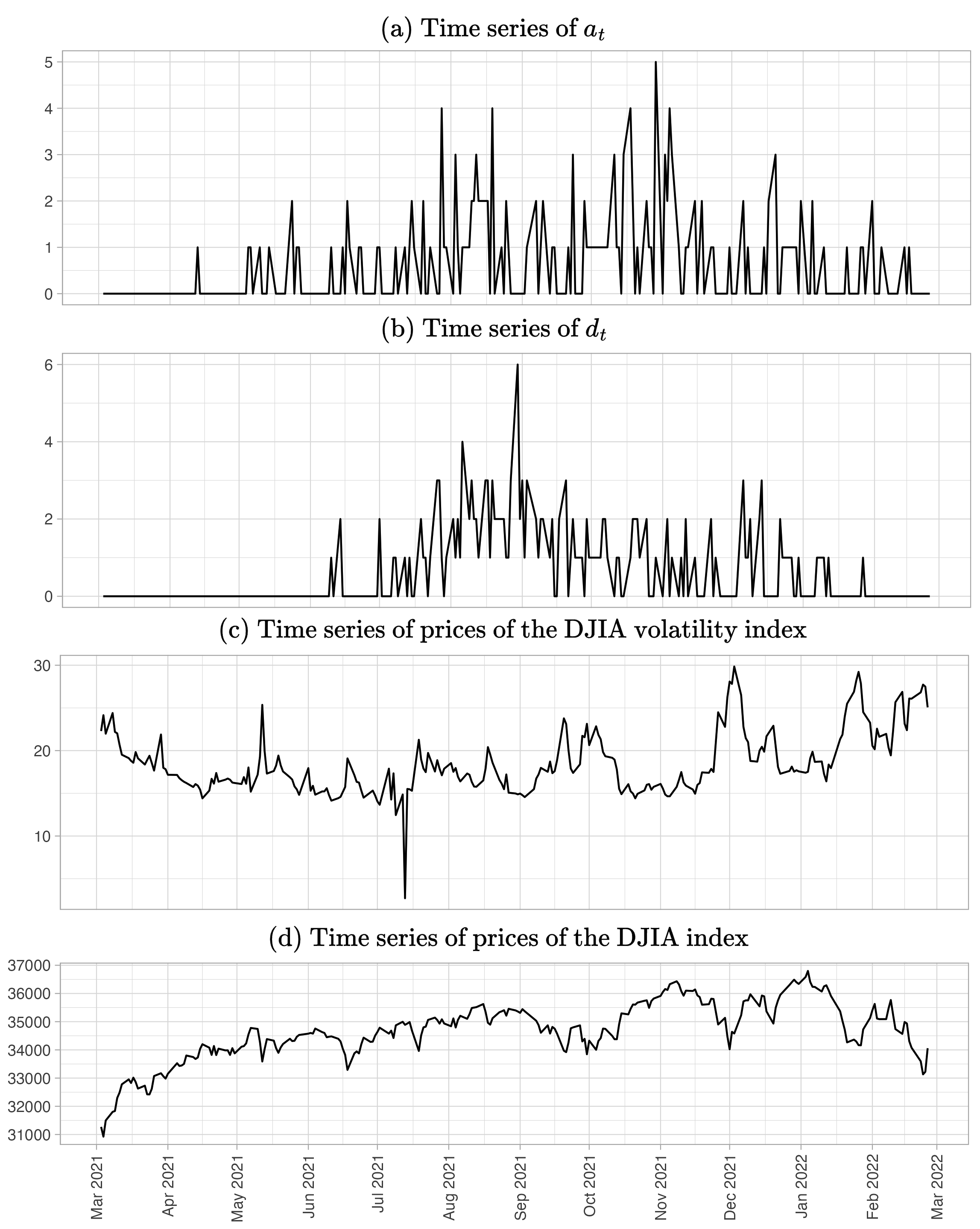}
	\caption{The times series of $a_t$, $d_t$, the DJIA Volatility Index, and DJIA prices in the first window.}
	\label{fig:DBN_AD_five_summary}
\end{figure}

\autoref{table:illustrating_first_window_DBN_AD} shows the parameter estimation results, and \autoref{fig:DBN_AD_five_summary} shows the times series of $a_t$, $d_t$, and network density. The time series of the DJIA Volatility Index and DJIA price are also included for reference. Note that $a^C+b^C=0.893$ is large, suggesting that the correlation dynamics are persistent. The parameters $\alpha_1+\beta_1=0.994$ and $\alpha_2+\beta_2=0.994$ are also large, suggesting that the dynamics of $a_t$ and $d_t$ are also persistent. The time series plots for $a_t$ and $d_t$ in \autoref{fig:DBN_AD_five_summary} show clustering patterns, which are typical patterns of persistent time series. Furthermore, $\gamma_1$ and $\gamma_2$ have the same sign and are both negative, suggesting that $a_t$ and $d_t$ are positively correlated, and they are negatively correlated with the volatility index in the first window as shown in \autoref{fig:DBN_AD_five_summary}. This can be explained by the increasing DJIA index prices in \autoref{fig:DBN_AD_five_summary}. First note that the volatility index and the DJIA index prices are negatively correlated with correlation coefficient -0.259. On the other hand, $a_t$ and $d_t$ are positively correlated with the DJIA index prices, with correlation coefficients respectively 0.30 and 0.29. These results suggest that the market is stably growing, and on average, the networks change more rapidly whenever the stock prices are high. This can be a sign of a bull market. From this example, including the DJIA prices in the dynamics for $a_t$ and $d_t$ may improve the model.

\begin{figure}[H]
	\centering
	\includegraphics[width=12cm]{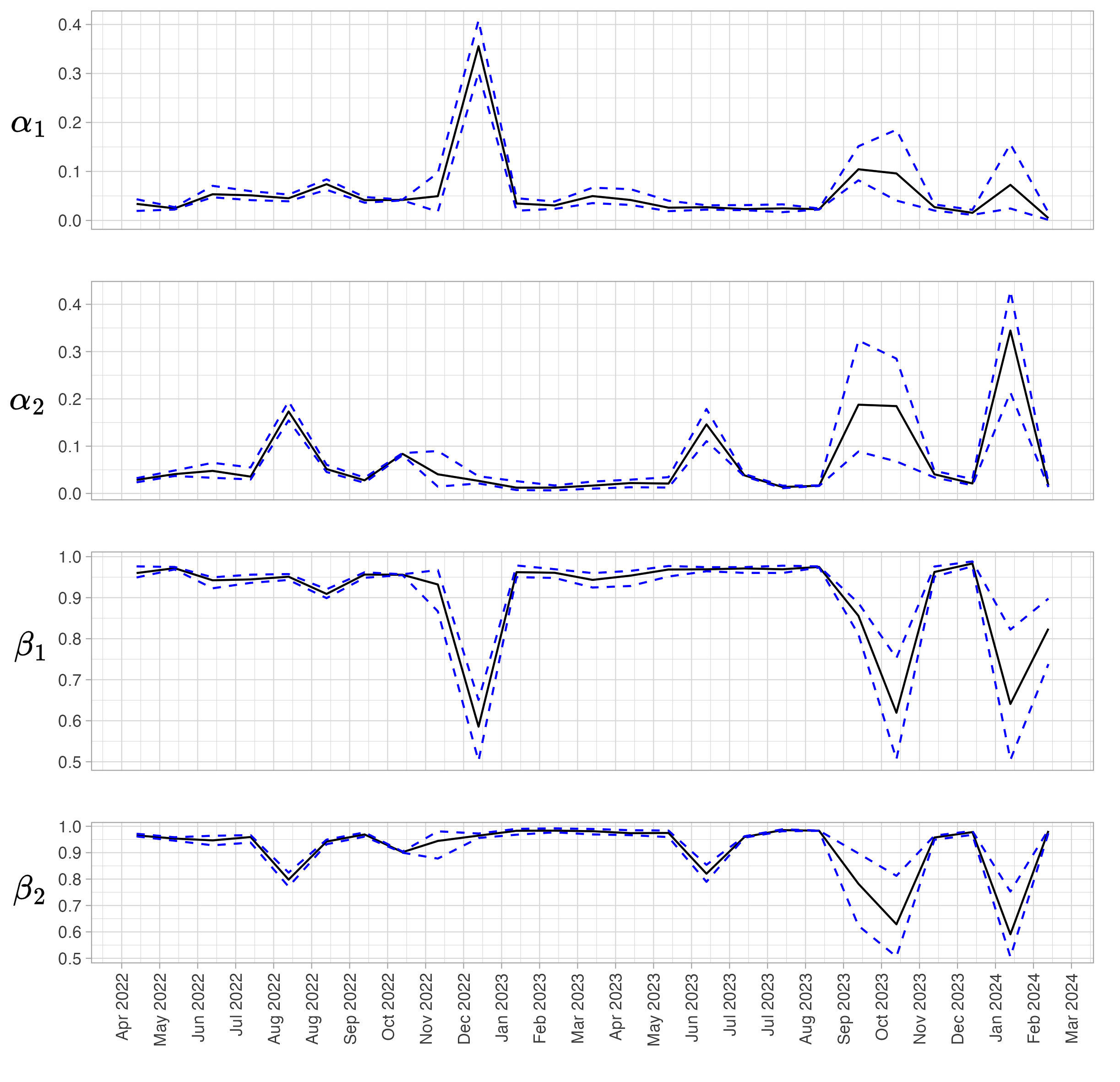}
	\caption{The times series of $\alpha_1$, $\alpha_2$, $\beta_1$, and $\beta_2$. The black solid lines indicate the posterior mean, and the pair of blue dotted lines indicate the 95\% credible intervals.}
	\label{fig:DBN_AD_alpha_beta_ts}
\end{figure}

\begin{figure}[H]
	\centering
	\includegraphics[width=12cm]{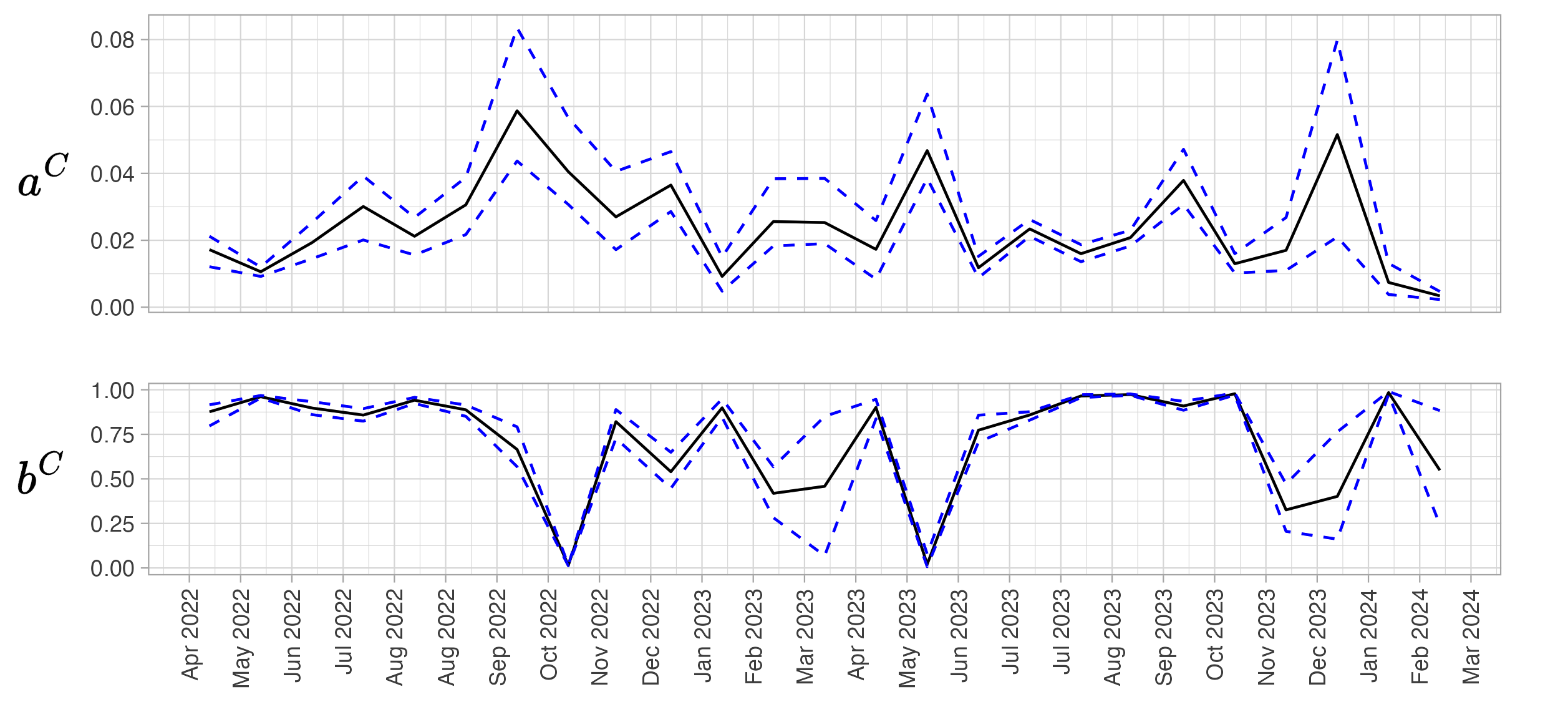}
	\caption{The times series of $a^C$ and $b^C$. The black solid lines indicate the posterior mean, and the pair of blue dotted lines indicate the 95\% credible intervals.}
	\label{fig:DBN_AD_aCbC_ts}
\end{figure}

\begin{figure}[H]
	\centering
	\includegraphics[width=12cm]{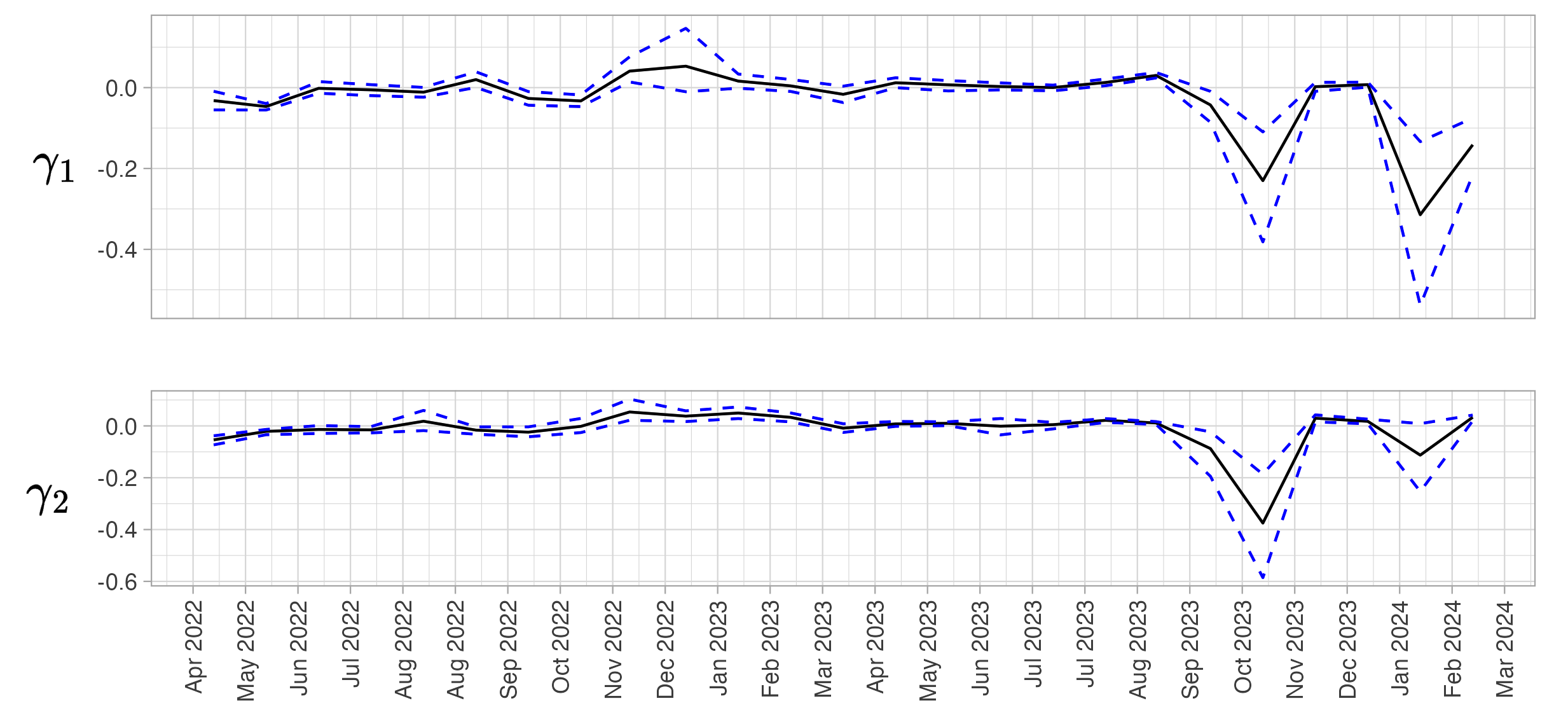}
	\caption{The times series of $\gamma_1$ and $\gamma_2$. The black solid lines indicate the posterior mean, and the pair of blue dotted lines indicate the 95\% credible intervals.}
	\label{fig:DBN_AD_gamma_ts}
\end{figure}

We also present the parameters in the fitted 25 models. \autoref{fig:DBN_AD_alpha_beta_ts} shows the times series of $\alpha_1$, $\alpha_2$, $\beta_1$, and $\beta_2$, \autoref{fig:DBN_AD_aCbC_ts} shows the time series of $a^C$ and $b^C$, and \autoref{fig:DBN_AD_gamma_ts} shows the times series of $\gamma_1$ and $\gamma_2$. In \autoref{fig:DBN_AD_aCbC_ts}, we observe that $b^C$ are large most of the time, and in \autoref{fig:DBN_AD_alpha_beta_ts}, $\beta_1$ and $\beta_2$ are close to 1 most of the time. These suggest that the conditional correlation dynamics and dynamics of $a_t$ and $d_t$ are persistent in most fitted models. The estimated parameters $\gamma_1$ and $\gamma_2$ shown in \autoref{fig:DBN_AD_aCbC_ts} have similar trends, suggesting that the volatility index drives the dynamics of $a_t$ and $d_t$ in a similar way, and $a_t$ and $d_t$ are positive correlated in most fitted models.

\begin{table}[H]
	\centering
	\begin{tabular}{lll}
		Name                    & Symbol & Market capitalization \\ \hline
		Microsoft               & MSFT   & \$3.029 T   \\
		Apple                   & AAPL   & \$2.801 T   \\
		Amazon                  & AMZN   & \$1.798 T   \\
		Visa                    & V      & \$586.70 B  \\
		JPMorgan Chase          & JPM    & \$531.08 B  \\
		Walmart                 & WMT    & \$480.10 B  \\
		UnitedHealth            & UNH    & \$460.87 B  \\
		Johnson \& Johnson       & JNJ    & \$389.13 B  \\
		Procter \& Gamble        & PG     & \$376.60 B  \\
		Home Depot              & HD     & \$375.82 B  \\
		Merck                   & MRK    & \$324.65 B  \\
		Salesforce              & CRM    & \$290.17 B  \\
		Chevron                 & CVX    & \$282.93 B  \\
		Coca-Cola               & KO     & \$260.47 B  \\
		McDonald                & MCD    & \$213.14 B  \\
		Walt Disney             & DIS    & \$203.24 B  \\
		Cisco                   & CSCO   & \$194.60 B  \\
		Intel                   & INTC   & \$177.53 B  \\
		IBM                     & IBM    & \$169.87 B  \\
		Verizon                 & VZ     & \$168.59 B  \\
	\end{tabular}
	\caption{$20$ stocks included in the empirical study.}
	\label{tab:stock_DBN}
\end{table}

\begin{table}[H]
	\centering
	\begin{tabular}{llll}
		Parameter     & Posterior mean & Lower 2.5\% & Upper 2.5\% \\ \hline
		$a^C$         & 0.017          & 0.012      & 0.021       \\
		$b^C$         & 0.876          & 0.797      & 0.916       \\
		$\bar{\mu}^a$ & 0.002          & 0.000  & 0.005       \\
		$\bar{\mu}^d$ & 0.001          & 0.000    & 0.004       \\
		$\beta^{ES}$  & 0.209          & 0.182      & 0.236       \\
		$\alpha_1$    & 0.034          & 0.019      & 0.043       \\
		$\alpha_2$    & 0.029          & 0.024      & 0.032       \\
		$\beta_1$     & 0.960          & 0.949      & 0.976       \\
		$\beta_2$     & 0.965          & 0.962      & 0.972       \\
		$\gamma_1$    & -0.032         & -0.055     & -0.009      \\
		$\gamma_2$    & -0.054         & -0.073    & -0.038     
	\end{tabular}
	\caption{The posterior means and the 95\% credible intervals of the sampled parameters in the first fitted model.}
	\label{table:illustrating_first_window_DBN_AD}
\end{table}

\subsection{Prediction}
In each window of $T=250$ days of observation, we conduct MCMC sampling to estimate the parameters. The posterior means of the MCMC sampled parameters and the maximum a posteriori (MAP) network on the last day in the data set are used in the DBN-AD model for prediction. To predict the correlation matrices, we adopt the Monte Carlo simulation prediction method. We generate 100 time series of networks and returns from the fitted DBN-AD model. For each time $t$ for $t=T,\ldots,T+T'-1$, where $T'$ is the number of trading days in the future four weeks, we do the following.

\begin{enumerate}
	\item 	We first generate $a_{t+1}$ and $d_{t+1}$ from  \eqref{eqt:dynamic_mean_spec}. For $t=T$, $\mu_{T}^a$ and $\mu_{T}^d$ are substituted by their posterior means in the MCMC sample, while we use $a_T$ and $d_T$ corresponding to the MAP network, as we need integer inputs.
	\item We estimate $\{w_{i,T}\}_{i=1}^n$ using the posterior means of the samples at time $T$. The individual activeness $\{w_{i,t+1}\}_{i=1}^n$ are calculated recursively using \eqref{eqt:dynamics_wit_reference_group}, and thus $G_{t+1}$  can be determined.
	\item The one-day ahead predicted correlation matrix is calculated using \eqref{eqt:cond_corr} according to the dependence structure of the predicted $G_{t+1}$. There are two ways to set $\rho_{i,i_t[k]|i_t[k-1],\ldots,i_t[1]}^{[T+1]}$: (1) Substituting the posterior mean of\\ $\rho_{i,i_t[k]|i_t[k-1],\ldots,i_t[1]}^{[T]}$ in the MCMC sample into  \eqref{eqt:cond_corr} to compute $\rho_{i,i_t[k]|i_t[k-1],\ldots,i_t[1]}^{[T+1]}$, or (2) Use the posterior mean of  $\overline\rho_{i,i_t[k]|i_t[k-1],\ldots,i_t[1]}$ as $\rho_{i,i_t[k]|i_t[k-1],\ldots,i_t[1]}^{[T+1]}$. While method (1) uses Bayesian averaging, the MCMC samples of $G_T$ may be evolved from different initial networks $G_1$ and different edge change dynamics in a horizon of 250 days. Thus, $G_T$ could be substantially varying among the samples, leading to large variances for the correlation matrix estimation at time $T$.  Hence, we use method (2) for the prediction. Then, we calculate $\rho_{i,i_t[k]|i_t[k-1],\ldots,i_t[1]}^{[t+1]}$ recursively using \eqref{eqt:cond_corr}.
	\item We generate conditional variances at time $t+1$ from the GARCH model in \eqref{eqt:GARCH} for all stocks using the posterior means of the MCMC sampled GARCH parameters. Then, we can calculate the conditional covariance matrix using these conditional variances and the conditional correlation matrix.
	\item A time series of returns is then generated from the multivariate normal distribution with a mean of zero and the conditional covariance matrix obtained in the previous step.
\end{enumerate}
For each $t=T+1,\ldots,T+T'$, the 100 Monte Carlo simulated covariance matrices at time $t$ are averaged out to get $\hat\Sigma_{t}$, the estimate of the covariance matrix at time $t$. The portfolio weights $\boldsymbol{\omega}_{t}=(\omega_{1,t},\ldots,\omega_{n,t})^T$ is determined by the minimum variance (MV) problem: $\min_{\boldsymbol{\omega}_{t}} \boldsymbol{\omega}_{t}^T\hat\Sigma_{t}\boldsymbol{\omega}_{t}$ such that $\sum_{i=1}^{n} \omega_{i,t} = 1$.

\subsection{Investment Experiments}

We will compare our proposed DBN-AD model to the DCC-GARCH model for a hypothetical investment study. Starting from a wealth of \$1,000, we invest in the 20 stocks. The portfolio weights are adjusted daily, by minimizing the portfolio variance calculated from the one-day ahead predicted covariance matrix using the DBN-AD model and the DCC-GARCH model. Once the model is updated, we predict the covariance matrices for the trading days in the future four weeks. The investment strategies are as follows.

\begin{enumerate}
	\item[(Strategy 1)] Invest on every trading day.
	\item[(Strategy 2)] Invest only when the one-day ahead predicted risk indicator is smaller than the average risk indicators on the previous $M$ trading days. Examples of risk indicators are the average network density, global clustering coefficients of the one-day ahead predicted network, and the one-day ahead predicted Value-at-Risk. Mathematically, let $s_{i,t}$ be the risk indicator at time $t$. On day $t \leq T+M$, we still invest every day. On day $t>T+M$, we invest into the financial market only when ${s}_{i,t}<\overline{s}_{i,t-1}:=\sum_{\tau=t-M}^{t-1}s_{i,\tau}/M$. We calculate the risk indicators as follows.
	\begin{enumerate}[(i)]
		\item 	Let $G_{t+1}^{(\ell)}$ be the $\ell$th Monte Carlo simulated network at time $t+1$, and let $ND_{t+1}^{(\ell)}$ and $CC_{t+1}^{(\ell)}$ be the network density and the clustering coefficient of $G_{t+1}^{(\ell)}$. The estimated average network density and global clustering coefficient at time $t+1$ are respectively given by $\overline{ND}_{t+1}=\sum_{\ell=1}^{100} ND_{t+1}^{(\ell)}/100$ and $\overline{CC}_{t+1}=\sum_{\ell=1}^{100}CC_{t+1}^{(\ell)}/100$.
		\item Let $\hat{\mathbf{r}}_{t+1}^{(\ell)}$ be the prediction of the returns of the $n$ stocks at time $t+1$ in the $\ell$th Monte Carlo sample. We first calculate the optimal portfolio weight $\boldsymbol{\omega}_{t+1}^{(\ell)}$ from the one-day ahead predicted covariance matrix in the $\ell$th Monte Carlo sample. Then, we calculate 100 portfolio returns $\mathcal{R}=\{(\boldsymbol{\omega}_{t+1}^{(\ell)})^T\hat{\mathbf{r}}_{t+1}^{(\ell)}:\ell=1,\ldots,100\}$ and the VaR at $100\alpha\%$ on day $t+1$ is predicted by the negative of the $\alpha$th quantile of $\mathcal{R}$.
	\end{enumerate}
	
\end{enumerate}

\begin{figure}[H]
	\centering
	\includegraphics[width=13cm]{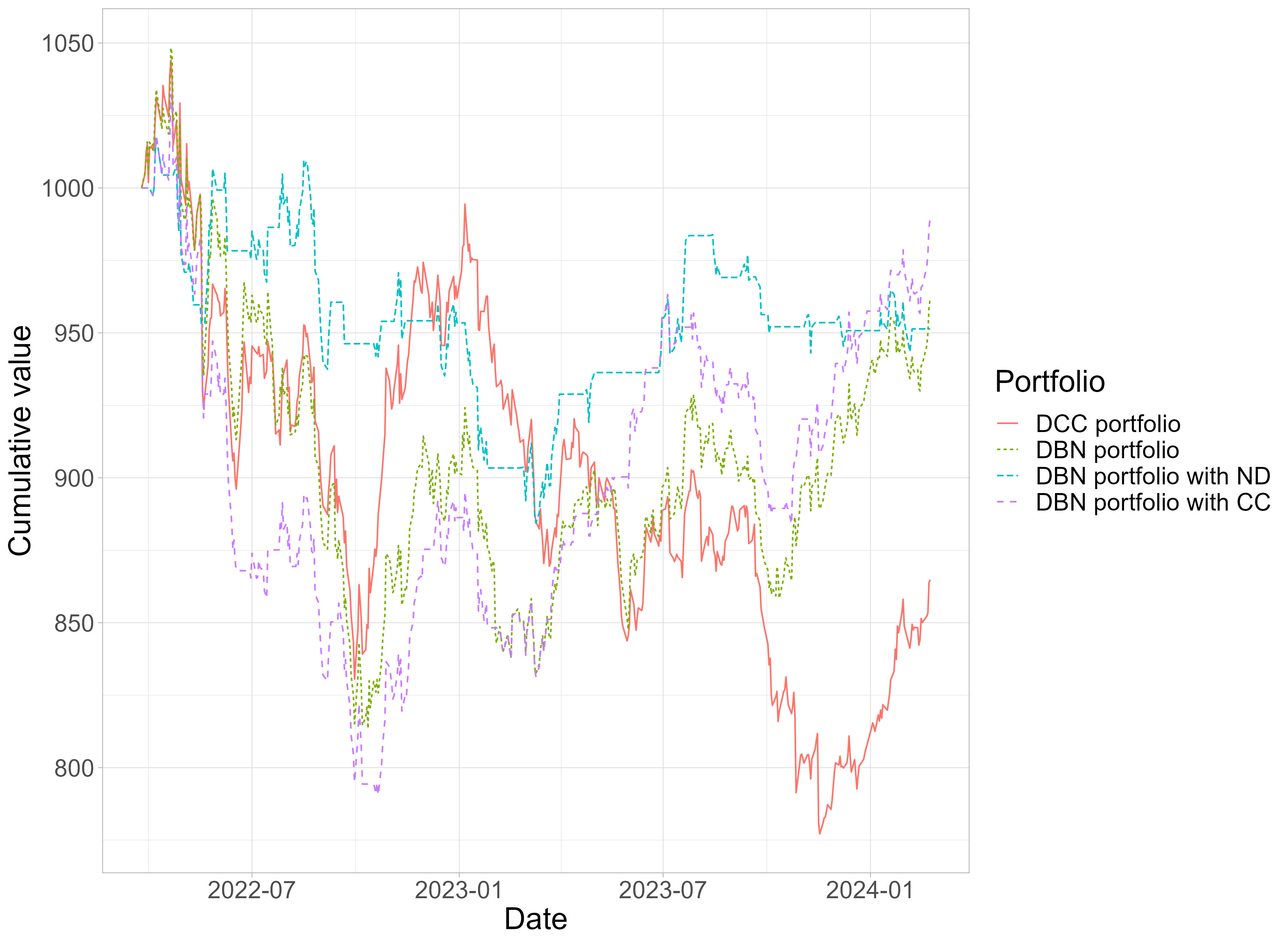}
	\caption{The cumulative values of the four portfolios: (1) DCC-GARCH model with strategy 1 (DCC portfolio), (2) DBN-AD model with strategy 1 (DBN portfolio), (3) DBN-AD model with strategy 2 using one-day ahead predicted network density as the risk indicator (DBN portfolio with ND), and (4) DBN-AD model with strategy 2 using one-day ahead predicted global clustering coefficient as the risk indicator (DBN portfolio with CC).}
	\label{fig:invest_original}
\end{figure}

\autoref{fig:invest_original} shows the cumulative values of the four portfolios: (1) DCC-GARCH model with strategy 1 (DCC portfolio), (2) DBN-AD model with strategy 1 (DBN portfolio), (3) DBN-AD model with strategy 2 using one-day ahead predicted network density as the risk indicator (DBN portfolio with ND), and (4) DBN-AD model with strategy 2 using one-day ahead predicted global clustering coefficient as the risk indicator (DBN portfolio with CC). We observe that the portfolios from the DBN-AD model perform better than the DCC portfolio in general, especially for portfolio (3), showing that predicting the network density from the DBN model is useful to avoid risk. Note that these risk indicators can only be obtained from the DBN-AD model but not the DCC-GARCH model.

\begin{figure}[H]
	\centering
	\includegraphics[width=13cm]{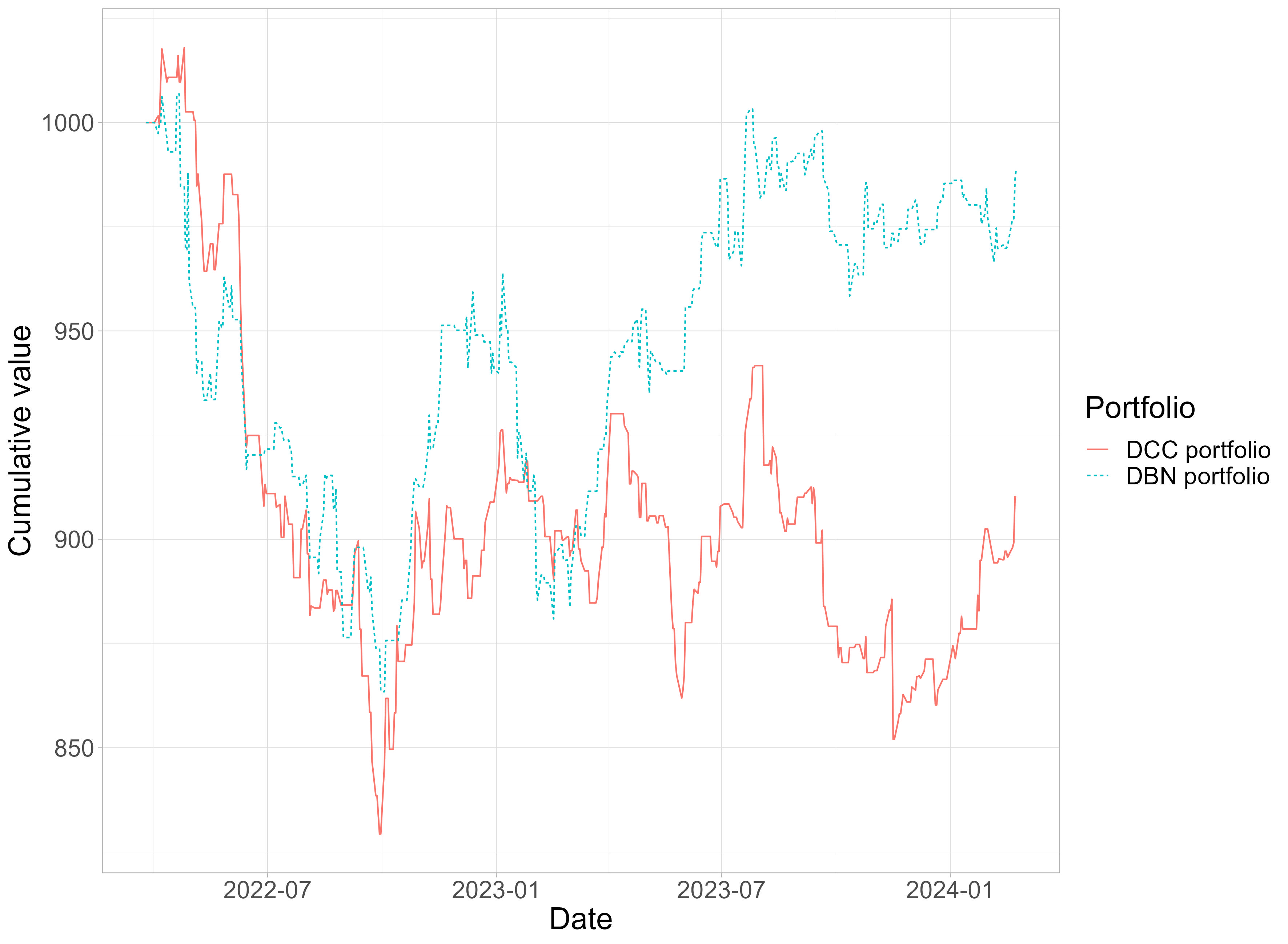}
	\caption{The cumulative values of the portfolios: (5) DCC-GARCH model with strategy 2 using one-day ahead predicted 2.5\% VaR as the risk indicator (DCC portfolio), and (6) DBN-AD model with strategy 2 using one-day ahead predicted 2.5\% VaR as the risk indicator (DBN portfolio).}
	\label{fig:invest_VaR25}
\end{figure}

\begin{figure}[H]
	\centering
	\includegraphics[width=13cm]{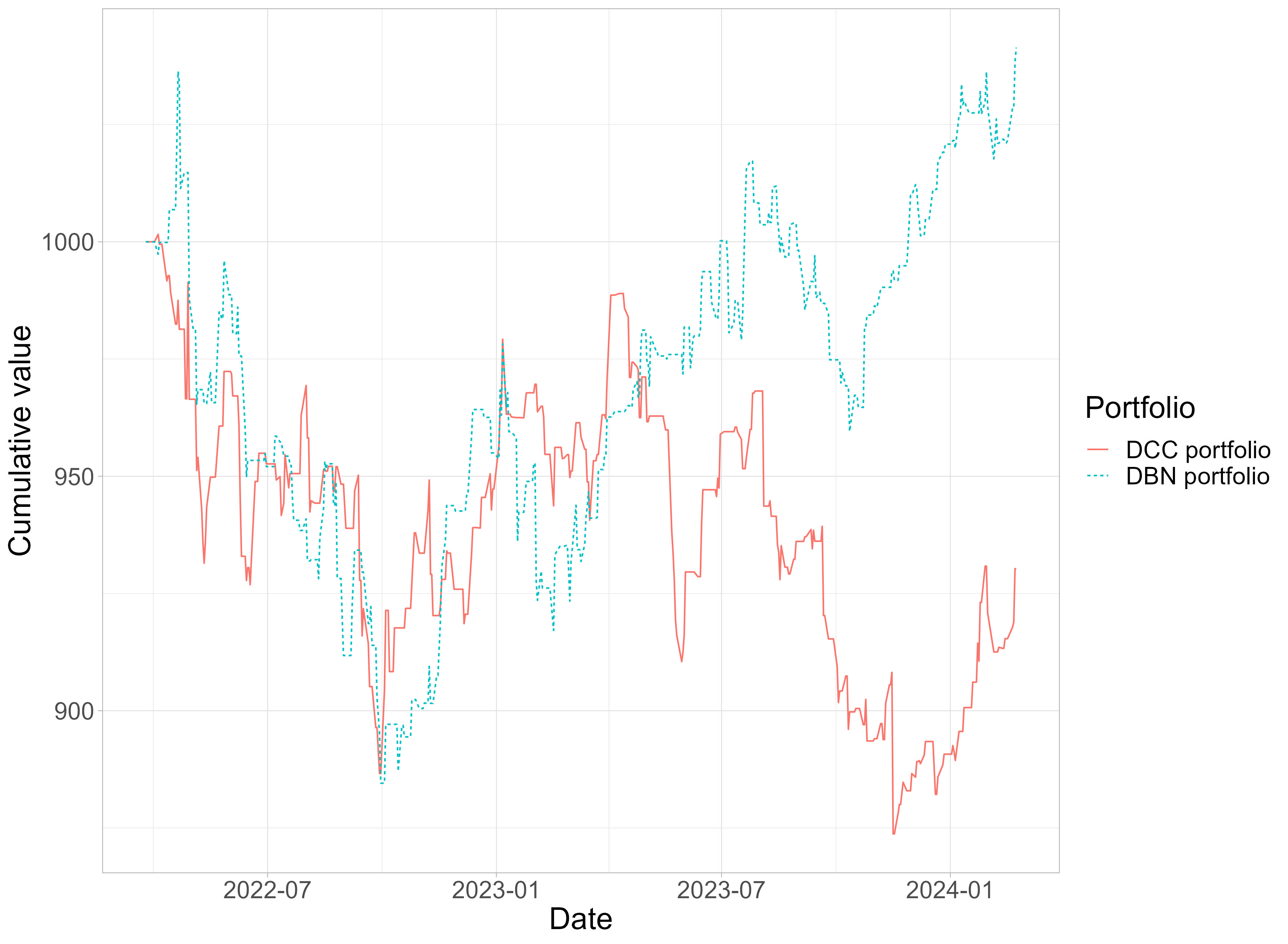}
	\caption{The cumulative values of the portfolios: (7) DCC-GARCH model with strategy 2 using one-day ahead predicted 1\% VaR as the risk indicator (DCC portfolio), and (8) DBN-AD model with strategy 2 using one-day ahead predicted 1\% VaR as the risk indicator (DBN portfolio).}
	\label{fig:invest_VaR1}
\end{figure}

To have a fairer comparison between the DCC-GARCH and DBN-AD models, we consider strategy 2 using Value-at-Risk as the risk indicator, where we can predict VaRs using both DCC-GARCH and DBN-AD models.  \autoref{fig:invest_VaR25} show the cumulative values of two more portfolios: (5) DCC-GARCH model with strategy 2 using one-day ahead predicted 2.5\% VaR as the risk indicator (DCC portfolio), and (6) DBN-AD model with strategy 2 using one-day ahead predicted 2.5\% VaR as the risk indicator (DBN portfolio). \autoref{fig:invest_VaR1} contains two portfolios similar to (5) and (6) except we use 1\% VaR as the risk indicator instead. The DBN portfolios in both strategies perform better. This suggests that the DBN-AD model has a better ability to detect risk.

\section{Discussion and Conclusion}
\label{section:conclusion}
In contrast to the existing models that often use a network mixture idea, which limits the frequency of the network changes or relies on structural priors to smooth the time series of networks, resulting in unknown evolution, our proposed model allows for gradual changes in networks over time in an intuitive way. We illustrate our model by considering a financial case using GARCH with normal distributed innovations. We allow the financial network to change gradually over time, mimicking the dynamic nature of the financial markets, where the edges in a BN can be thought of as the direction of the propagation of volatility turbulence. 

The simulation study suggests that the parameters in the model can be accurately estimated, and the networks near the right end time points can be estimated with good accuracy. The accuracy can be further improved if we allow the MCMC sampler to run for longer, whereas we limit the estimation time to around 2 days to align with the real situation that we update the model at weekends. We allow some inaccuracies for the initial parts of the dynamic networks, which are not used in the prediction, in exchange for the shorter computational time. We tailor-make an MCMC estimation methodology for our proposed model. The initialization using a moving-window approach in Section \ref{section:network_initialization} provides a good initial time series of networks for the MCMC sampling. The block sampling methods using a random-walk move and a cyclic move introduced in Section \ref{section:sampling_for_at_dt} for $a_t$ and $d_t$ also provide better mixing than individual updates. 

The empirical study shows that the portfolios built using our proposed model provide better investment performances than those using the traditional DCC-GARCH model.  By setting up investment strategies using the risk indicators derived from the dynamic networks, we can further improve investment performances. We observe that the investment performances using the VaR estimated from our proposed model as risk indicators are much better than using the VaR estimated from the DCC-GARCH model, suggesting that our model has a better ability to detect risk by incorporating dynamic features in the dependence structure.

Although we demonstrate our model using financial data, it is possible to extend our model for application to other disciplines. The most important idea of our model is the intuitive dynamics in edge changes, while the GARCH and dynamic correlation specifications are not necessary. Furthermore, the current setting imposes strong restrictions on the trajectory of the edge changes, where only the edges with the largest and smallest activeness are added and removed. It is also possible to give further flexibility by allowing the addition and deletion lists to skip some edges. We left this extension for future research.

\section*{Acknowledgement} 

This work was supported by the Hong Kong RGC General Research Fund (grant number 16507322).

\bibliography{references}

\end{document}